\documentclass[preprint,showpacs,preprintnumbers,amsmath,amssymb,floatfix]{revtex4}
\usepackage{graphicx,color}% Include figure files
\usepackage{dcolumn}% Align table columns on decimal point
\usepackage{bm}% bold math
\usepackage{supertabular}

\begin{document}

%Title of paper
%\title{The atmospheric muon and neutrino fluxed II}
\title{Calculation of atmospheric neutrino flux using the interaction model calibrated with atmospheric muon data}
\author{M.~Honda}
\email[]{mhonda@icrr.u-tokyo.ac.jp}
\homepage[]{http://icrr.u-tokyo.ac.jp/~mhonda}
\author{T.~Kajita}
\email[]{kajita@icrr.u-tokyo.ac.jp}
\affiliation{Institute for Cosmic Ray Research, the University of Tokyo, 5-1-5 Kashiwa-no-ha, Kashiwa, Chiba 277-8582, Japan}
\author{K.~Kasahara}
\email[]{kasahara@icrc.u-tokyo.ac.jp}
\affiliation{Shibaura Institute of Technology, 307 Fukasaku, Minuma-ku, Saitama 330-8570, Japan}
\author{S.~Midorikawa}
\email[]{midori@aomori-u.ac.jp}
\affiliation{Faculty of Software and Information Technology, Aomori University, 2-3-1 Kobata, Aomori 030-0943, Japan.}
\author{T.~Sanuki}
\email[]{sanuki@icepp.s.u-tokyo.ac.jp}
\affiliation{International Center for Elementary Particle Physics, the University of Tokyo, 7-3-1 Hongo, Bunkyo-ku, Tokyo 113-0033, Japan}

\date{\today}

\begin{abstract}
Using the ``modified DPMJET-III'' model explained 
in the previous paper~\cite{shkkm2006}, 
we calculate the atmospheric neutrino flux.
The calculation scheme is almost the same as HKKM04~\cite{HKKM2004},
but the usage of the ``virtual detector'' is improved to reduce the
error due to it.
Then we study the uncertainty of the calculated atmospheric neutrino
flux summarizing the uncertainties of individual components of the simulation.
The uncertainty of $K$-production in the interaction model
is estimated using other interaction models: FLUKA'97 and Fritiof 7.02, 
and modifying them
so that they also reproduce the atmospheric muon flux data correctly.
The uncertainties of the flux ratio and zenith angle dependence of 
the atmospheric neutrino flux are also studied.
\end{abstract}

% insert suggested PACS numbers in braces on next line
\maketitle

 \section{\label{sec:introduction} Introduction}

In the previous paper~\cite{shkkm2006} (hereafter Paper~I),
we have studied the interaction model (DPMJET-III~\cite{Roesler:2000he}) 
employed in the
HKKM04 atmospheric neutrino flux calculation \cite{HKKM2004},
using the atmospheric muon flux data observed by 
precision measurements \cite{BESSTeVpHemu,BESSnorimu,l3+c}.
In the study, 
we found that the calculated and observed muon fluxes
did not agree, especially for momenta above 30~GeV/c.
We modified the interaction model to improve the agreement 
between the calculated and observed atmospheric muon fluxes.
We call this modified interaction model the modified DPMJET-III
in this paper.
Note, the modification is actually applied to the 
``inclusive DPMJET-III'' in a phenomenological way
based on the quark-parton model.

In this paper, we calculate the atmospheric neutrino flux
with the modified DPMJET-III (Sec.~\ref{sec:calculation}).
The calculation scheme and the physical input data are basically 
the same as the HKKM04.
However, we update the geomagnetic model from IGRF2000 to IGRF2005~\cite{igrf},
and improve the use of ``virtual detector'' in the 3-dimensional 
calculation to reduce the error due to it~\cite{barr2004}.

There are other hadronic interaction models which are used 
in the detector simulations of high energy experiments,
such as FLUKA'97~\cite{FLUKA97} 
and Fritiof 7.02~\cite{Pi:1992ug}.
We calculate the atmospheric neutrino fluxes with these
interaction models applying the modification, so that they
also reproduce the atmospheric muon flux observed by the precision
measurements
%, as the modified DPMJET-III 
(Sec.~\ref{sec:others}).
Note, to reproduce the observed muon flux, 
modifying the primary flux model might be alternative
solution.
We calculated the atmospheric neutrino flux,
changing the spectral index of primary cosmic ray protons 
from $-$2.71 to $-$2.66 above 100~GeV, which also reproduces the
observed muon flux in $\mu^+ + \mu^-$ sum correctly with
the original DPMJET-III.

Those calculations give almost the same atmospheric neutrino flux
in the energy region below 100~GeV, where $\pi$'s are the main source 
of atmospheric neutrinos.
With the modifications based on the atmospheric muon data,
the $\pi$ production is almost the same in all three
calculations.
However the $K$'s are not related to the atmospheric muons below 1~TeV/c,
and above this momenta, almost no muon flux is available from the 
precision measurements.
There remain sizable differences in the $K$ production, resulting in 
differences in atmospheric neutrino fluxes at higher energies.

In Sec.~\ref{sec:systematic}, we estimate the uncertainties in our 
calculations.
As the uncertainties of the predicted atmospheric neutrino 
flux is crucial for the study of neutrino oscillations,
the study of them is important~\cite{barr2006}.
Since our calculation reproduces the observed atmospheric muon flux data,
the uncertainty due to that in the $\pi$ production 
could be estimated from the experimental error and the residual of the 
reconstruction of atmospheric muon data.
The uncertainty of $K$ production is estimated from the variation of 
the atmospheric neutrino flux at higher energies calculated in the 
modified calculation schemes. 
%The flux variation itself is considered as the uncertainty of 
%atmospheric neutrino flux due to the uncertainty of the $K$ production.
The total uncertainty is estimated by summarizing individual 
uncertainties.
Note, 
the stabilities of the $(\nu_\mu+\bar\nu_\mu)/(\nu_e+\bar\nu_e)$ ratio 
and the zenith angle dependence of the atmospheric neutrino flux
are especially important in the study of neutrino oscillations.
They are also studied with the uncertainty of the flux value.

\section{\label{sec:calculation} Calculation of Atmospheric Neutrino Flux}

In this section, we describe the calculation of the atmospheric neutrino flux
with the modified DPMJET-III in detail.
The calculation scheme 
%of the atmospheric neutrino flux
is basically the same as HKKM04~\cite{HKKM2004}.
For the primary flux model, we take the same primary flux model
as HKKM04, based on AMS \cite{AMS1p, AMS1He} and 
BESS \cite{BESSpHe,BESSTeVpHemu} data,
with a spectral index of $-$2.71 above 100~GeV
(see also Refs.\ \cite{Gaisser-hamburg,Gaisser-Honda}).
For the model of the atmosphere, we used the 
US-standard '76 \cite{us-standard}, 
as the error due to the atmospheric density model 
is sufficiently small for the calculation of atmospheric neutrino 
flux~\cite{shkkm2006}.
Note, however,
we update the geomagnetic field model from IGRF2000 to IGRF2005~\cite{igrf},
and improve of the usage of the ``virtual detector'' as explained in 
this section.

Note, there are a considerable number of 3-dimensional calculations 
of the atmospheric neutrino 
flux~\cite{hkkm-dipole,liu2002,battis2002,battis2003,wentz2003,favier2003,barr2004}.
However, all those calculations suffer from the small statistics at higher 
neutrino energies ($\gtrsim$~10 GeV), due to the inefficiency of the 
3-dimensional calculation scheme.
In this paper, we calculate the atmospheric neutrino flux 
averaged over all azimuthal angles,
combining 3-dimensional and 1-dimensional calculations.
In HKKM04, it is shown that the atmospheric neutrino flux calculated
with a 1-dimensional scheme agrees with that calculated 
with the 3-dimensional scheme above a few GeV, averaged over all
azimuthal angles,
although more than a few~\% %($\sim$~10 ~\%)
azimuth angle dependence remains in the atmospheric 
neutrino flux at 10~GeV due to muon curvature in the geomagnetic field.

For the 3-dimensional calculation,
we assume the surface of the Earth is a sphere with $R_e=6378.180$~km.
We also assume three more spheres; 
the injection, simulation, and escape spheres.
The radius of the injection sphere is taken as $R_{inj}= R_e + 100$~km,  
the simulation sphere as $R_{sim}= R_e + 3000$~km, 
and the escape sphere as  $R_{esc}= 10 \times$$R_e$.
Note, a spheroid with an eccentricity of $\sim$1$/$298 is a better 
approximation for the Earth.
However, the differece from the sphare approximation is small,
and we estimate the errors of the atmospheric neutrino flux
due to the sphere approximation is also small ($\lesssim$~1\%).

Cosmic rays are sampled on the injection sphere uniformly 
toward the inward direction, 
following the given primary cosmic ray spectra. 
Before they are fed to the simulation code for the propagation in air, 
they are tested to determine whether they pass the rigidity cutoff, i.e., 
the geomagnetic barrier.
For a sampled cosmic ray, the `history' is examined by solving the 
equation of motion in the negative time direction.
When the cosmic ray reaches the escape sphere without touching the injection
sphere again in the inverse direction of time, 
the cosmic ray can pass through the magnetic barrier following its trajectory 
in the normal direction of time.

The propagation of cosmic rays is simulated in the space between 
the surface of Earth and the simulation sphere.
When a particle enters the Earth, it loses
its energy very quickly, and results in 
neutrinos with energy less than 100~MeV.
Therefore, we discard such particles as soon as they enter the Earth, as
most neutrino detectors which observe atmospheric neutrinos do
not have sensitivity below 100~MeV.

For secondary particles produced in the interaction of a cosmic ray 
and air-nucleus, there is the possibility that they go out from 
and re-enter in the atmosphere and create low energy neutrinos.
Therefore, a simulation sphere which is too small 
may miss such secondary particles.
On the other hand, it is very time consuming to follow all 
particles out to distances far from the Earth.
In HKKM04, 
the simulation sphere with radius 
$R_{sim} = R_{e} + 3000$ km was found to be large enough
to suppress the error to well below the 1~\% level.

\begin{figure}[tbh]\centerline{
\includegraphics[width=7cm]{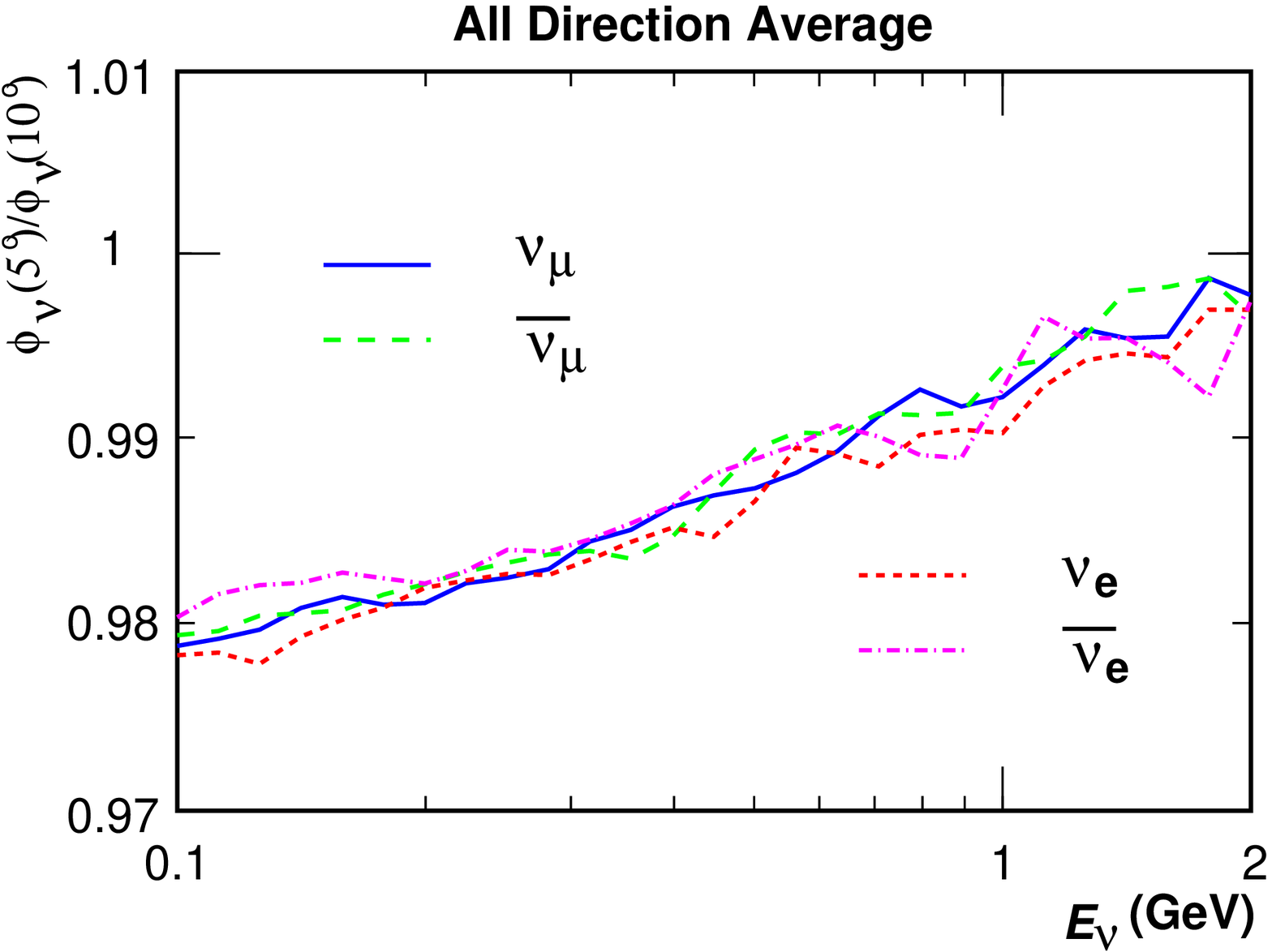}%
\hspace{5mm}
\includegraphics[width=7cm]{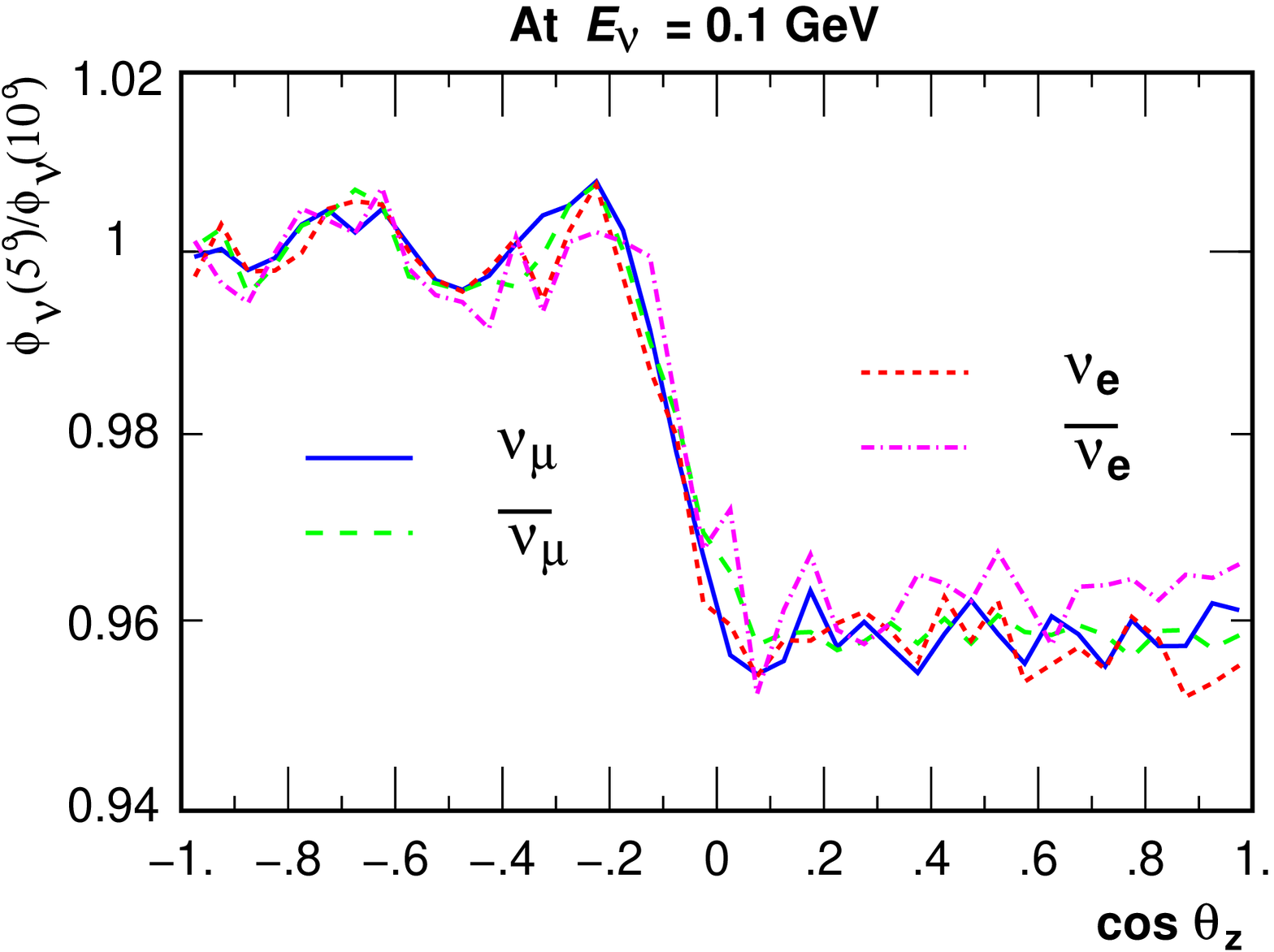}%
}\caption{\label{fig:f10-o-f05}
Left panel: Ratio of all direction averaged flux with a
smaller virtual detector ($\Phi_\nu(\theta_d=5^\circ)$) to that with the 
larger virtual detector ($\Phi_\nu(\theta_d=10^\circ)$) used in HKKM04.
Right panel: Zenith angle ($\theta_z$) dependence of the ratio at $E_\nu=0.1$~GeV
in azimuthal average.
}
\end{figure}

Note, 
neutrino detectors are very small 
compared with the size of the Earth, 
and are considered as the infinitesimal points on the 
surface of the Earth.
We introduce a finite size ``virtual detector'' 
for each targert detector in the 3-dimensional calculation scheme.
In HKKM04, the surface of the Earth within a circle
around the target detector of radius $\theta_d = 10^\circ$ 
($\sim$~1000~km) is used as the virtual detector.
When neutrinos pass throuth the surface of the Earth 
inside of the circle (upward or downward),
they are registered.
We do not need the virtual detector in 1-dimensional calculation 
scheme, 
since it treats the propagation of the cosmic rays on a line which
go through the neutrino detector.
This is a far less time consuming computation scheme than the
3-dimensional calculation scheme for the atmospheric neutrino flux, 

The finite size of the virtual detector introduces an error, 
since it averages the neutrino flux over positions where
the geomagnetic conditions are different from the position of
target detector~\cite{barr2004}.
To study the relation between the size and the error, 
we calculate the atmospheric neutrino flux with 
different size virtual detectors,
$\theta_d = 10^\circ$ ($\Phi_\nu(10^\circ)$) and 
$\theta_d=5^\circ $ ($\Phi_\nu(5^\circ)$).
The fluxes $\Phi_\nu(10^\circ)$ and $\Phi_\nu(5^\circ)$ are compared 
in Fig.~\ref{fig:f10-o-f05}
for Kamioka from the HKKM04 calculation averaging over
all azimuthal angles.
We find a difference is seen for downward directions,
and is almost constant for $\cos\theta_z > 0$ in ratio,
where $\theta_z$ is the zenith angle of the arrival direction
of the neutrinos.
Therefore we expect the maximum error at $E_\nu=0.1$~GeV is $\sim$~5~\% 
for downward directions averaging over azimuthal angles.
The error due to the finite size virtual detector 
are smaller than those due to uncertainty of hadronic interaction model
in HKKM04.

We can reduce the error due to the finite size virtual detector,
with a little more computation.
Let us assume the ``true'' atmospheric neutrino flux is expressed 
as an analytic function of the position.
Note, we drop the arguments for arrival direction in the following 
expressions. The discussion here should apply to each arrival 
direction independently.
We consider the power expansion the analytic function as,
\begin{equation}
\label{eq:analytic}
\phi_\nu(\theta_x, \theta_y) 
= \phi_\nu^{(0,0)} 
+ 
\phi_\nu^{(1,0)}\cdot \theta_x +
\phi_\nu^{(0,1)}\cdot \theta_y
+ 
\phi_\nu^{(2,0)}\cdot\theta_x^2 +
\phi_\nu^{(1,1)}\cdot\theta_x \theta_y +
\phi_\nu^{(0,2)}\cdot\theta_y^2
+ 
\cdots
%\end{array}
\end{equation}
and
\begin{equation}
\phi_\nu^{(m,n)} \equiv \frac{1}{ m! \cdot n!} \cdot 
\frac{\partial^{m+n} \phi_\nu}{\partial\theta_x^m\partial\theta_y^n} 
\Big|_{(\theta_x, \theta_y)=(0,0)}~,
\end{equation}
where, $\theta_x,\theta_y$ are the distances from the target detector
in center angle to any directions perpendicular to each other, say, 
to South and East respectively, i.e., ($\theta_x,\theta_y$) 
constitute a local coordinate system.

In the Monte Carlo calculation of the atmospheric neutrino flux, 
the calculated flux with a finite size virtual detector is the 
average flux over the virtual detector.
With the increase of statistics,
the flux $\Phi(\theta_d)$ calculated in Monte Carlo calculation 
should approach 
\begin{equation}
\label{eq:flux-average}
%\Phi_\nu(\theta_d) ~\simeq
\frac{1}{S(\theta_d)}
\int_{\theta_r < \theta_d} \phi_\nu(\theta_x,\theta_y) 
d\theta_x d\theta_y~~(\theta_d \ll 1),
\end{equation}
where $S(\theta_d)\simeq\pi\theta_d^2$ is the ``area'' of the virtual detector,
and $\theta_r\simeq\sqrt{\theta_x^2 + \theta_y^2}$.
The integrations of terms proportional to $\theta_x$ or $\theta_y$ 
in Eq.~\ref{eq:analytic} vanish,
and non-vanishing terms start from the integrations of
second order terms,
$
\phi_\nu^{(2,0)}\theta_x^2 +
\phi_\nu^{(1,1)}\theta_x \theta_y +
\phi_\nu^{(0,2)}\theta_y^2
$,
resulting in the terms  proportional to $\theta_d^4$.
For a sufficiently small $\theta_d$,
$\Phi_\nu(\theta_d)$ is expressed as,
\begin{equation}
\label{eq:conversion}
\Phi_\nu(\theta_d) 
\simeq \Phi^{(0,0)} + \frac{\Phi^{(2)}\theta_d^4}{S(\theta_d)} 
= \Phi^{(0,0)} + \Phi^{(2')}\theta_d^2~,
\end{equation}
where $\Phi^{(2')} \equiv \Phi^{(2)}\theta_d^2/{S(\theta_d)} \simeq \Phi^{(2)}/\pi$.
When we have the neutrino fluxes calculated with two virtual detectors
with small enough radii $\theta_d$ and $\theta_d/2$ for the same target, we expect 
$\Phi_\nu(\theta_d) - \Phi_\nu(\theta_d/2) \simeq \Phi^{(2')}\cdot [\theta_d^2 -(\theta_d/2)^2]$,
then $\Phi_\nu(0)$, the true flux value at the target detector, is given as
\begin{equation}
\label{eq:estimation}
\Phi_\nu(0)=\Phi^{(0,0)} \simeq \Phi_\nu(\theta_d) - 
\frac {4}{3}\cdot[\Phi_\nu(\theta_d) - \Phi_\nu(\theta_d/2)]~.
\end{equation} 

\begin{figure}[tbh] \centerline{
\includegraphics[width=7cm]{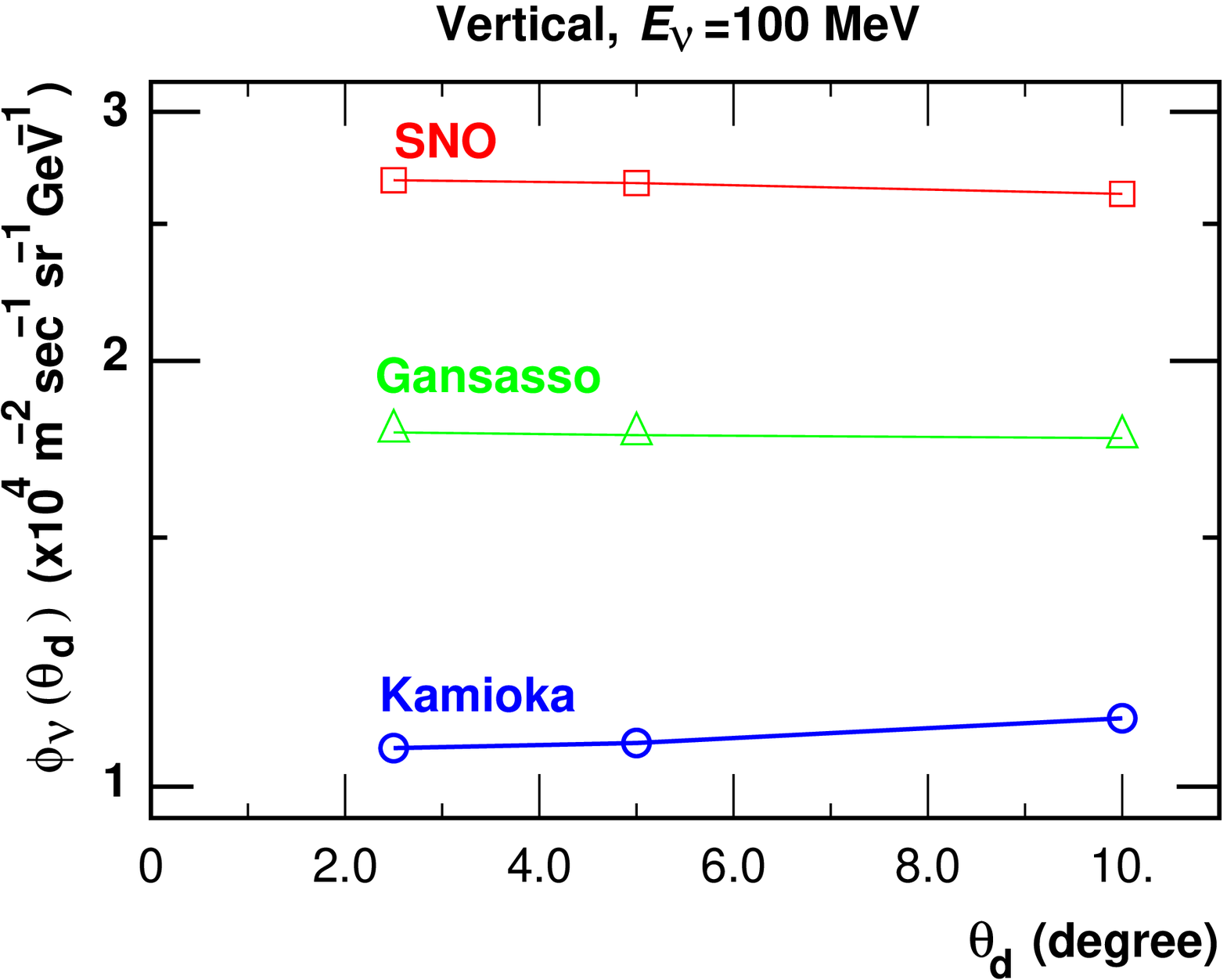}%
~~~~
\includegraphics[width=7cm]{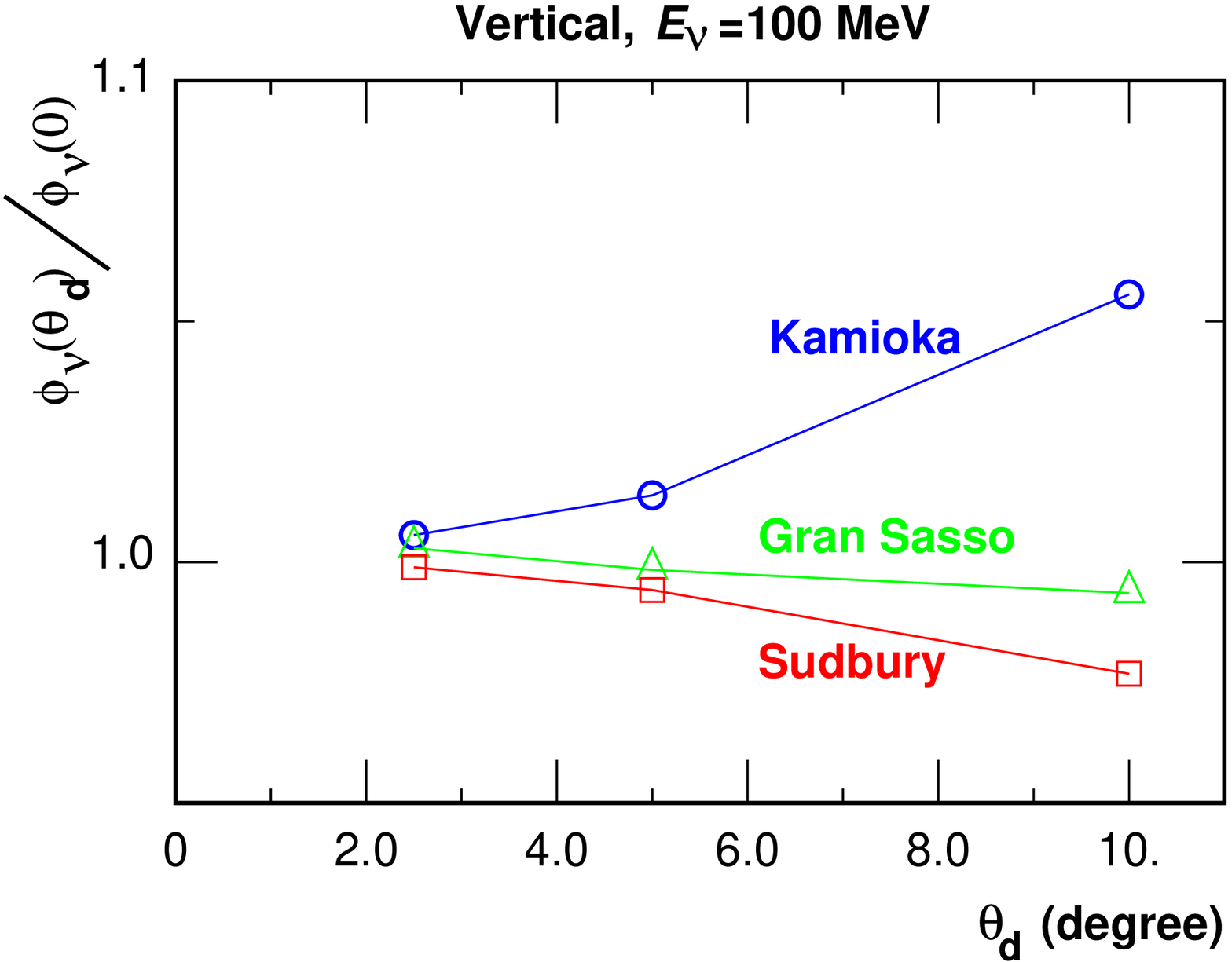}%
}
\caption{\label{fig:dep-vr}
Left: Atmospheric neutrino fluxes for vertical directions
calculated with the virtual detectors with different radii.
Right: Ratio of fluxes calculated with virtual detectors with 
$\theta_d=10, 5,$ and $2.5^\circ$ to the flux estimated 
with Eq.~\ref{eq:estimation}.
}
\end{figure}

As is seen in Fig.~\ref{fig:f10-o-f05}, the difference of the 
$\Phi_\nu(10^\circ)$ and $\Phi_\nu(5^\circ)$ 
%$\theta_d=10^\circ$ and $\theta_d=5^\circ$ 
is almost constant for $\cos\theta_z> 0$, 
so it should be sufficient 
%may be enough
to examine the assumption and procedure for vertical 
down going directions.
In the left panel of Fig.~\ref{fig:dep-vr},
we plotted the total neutrino flux for the vertical down going 
directions ($\cos\theta_z > 0.9$) calculated with different size of 
virtual detectors, $\theta_d=10^\circ, 5^\circ$, 
and $2.5^\circ$ %($\sim$ 279, 557, and 1117 km)
for Kamioka, Sudbury, and Gran Sasso with the HKKM04 calculation.
In the right panel of  Fig.~\ref{fig:dep-vr},
we depicted the difference to the estimated true value with
Eq.~\ref{eq:estimation} in the ratio.
We may say 
the convergence of the calculated fluxes to the ``true value'' agrees 
with the expectation of Eq.~\ref{eq:conversion},
and we apply the Eq.~\ref{eq:estimation} with $\theta_d=10^\circ$ and $5^\circ$ to 
the atmospheric neutrino flux calculated in the 3-dimensional scheme.
Note, the error due to the finite size of virtual detector does not 
exist in the 1-dimensional calculation scheme.

Thus calculated atmospheric neutrino fluxes in the 3-dimensional 
scheme are shown in the Appendix~A up to 10~GeV 
for Kamioka, Sudbury, and Gran Sasso separately.
The neutrino flux calculated for the Soudan2 site is almost
identical to that calculated for Sudbury.
The neutrino flux calculated for Frejus is $\sim$~10~\%
larger than that for Gran Sasso at 0.1~GeV,
and the difference is smaller at higher energies.
Above 10~GeV, the atmospheric neutrino flux is calculated 
using the 1-dimensional scheme. They are tabulated in the Appendix~B up to
10~TeV.

\begin{figure}[tbh] \centerline{
\includegraphics[width=5cm]{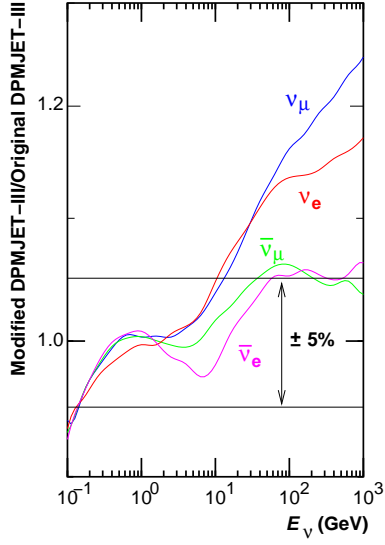}%
}
\caption{\label{fig:new-o-old}
The comparison of the atmospheric neutrino fluxes calculated
with modified and original DPMJET-III in ratio.
The denominator is the original DPMJET-III.
}
\end{figure}

We compared the atmospheric neutrino fluxes 
calculated with modified and original DPMJET-III in the ratio
of flux values in Fig~\ref{fig:new-o-old}, 
averaging over all directions for Kamioka up to 1~TeV.
The atmospheric neutrino flux calculated with the modified DPMJET-III
shows an increase above 10~GeV from that with the original DPMJET-III, 
but the increase rate is different for different kinds of neutrinos.
This is because 
the modification of the interaction model in Paper~I enhances the
productions of $\pi^\pm$'s, $K^+$'s, and $K^0$,
with no change for $K^-$ production.
Therefore, the increase of $\nu_\mu$ and $\nu_e$ is
larger than that of  $\bar\nu_\mu$ and $\bar\nu_e$.

\begin{figure}[tbh] \centerline{
\includegraphics[width=5.15cm]{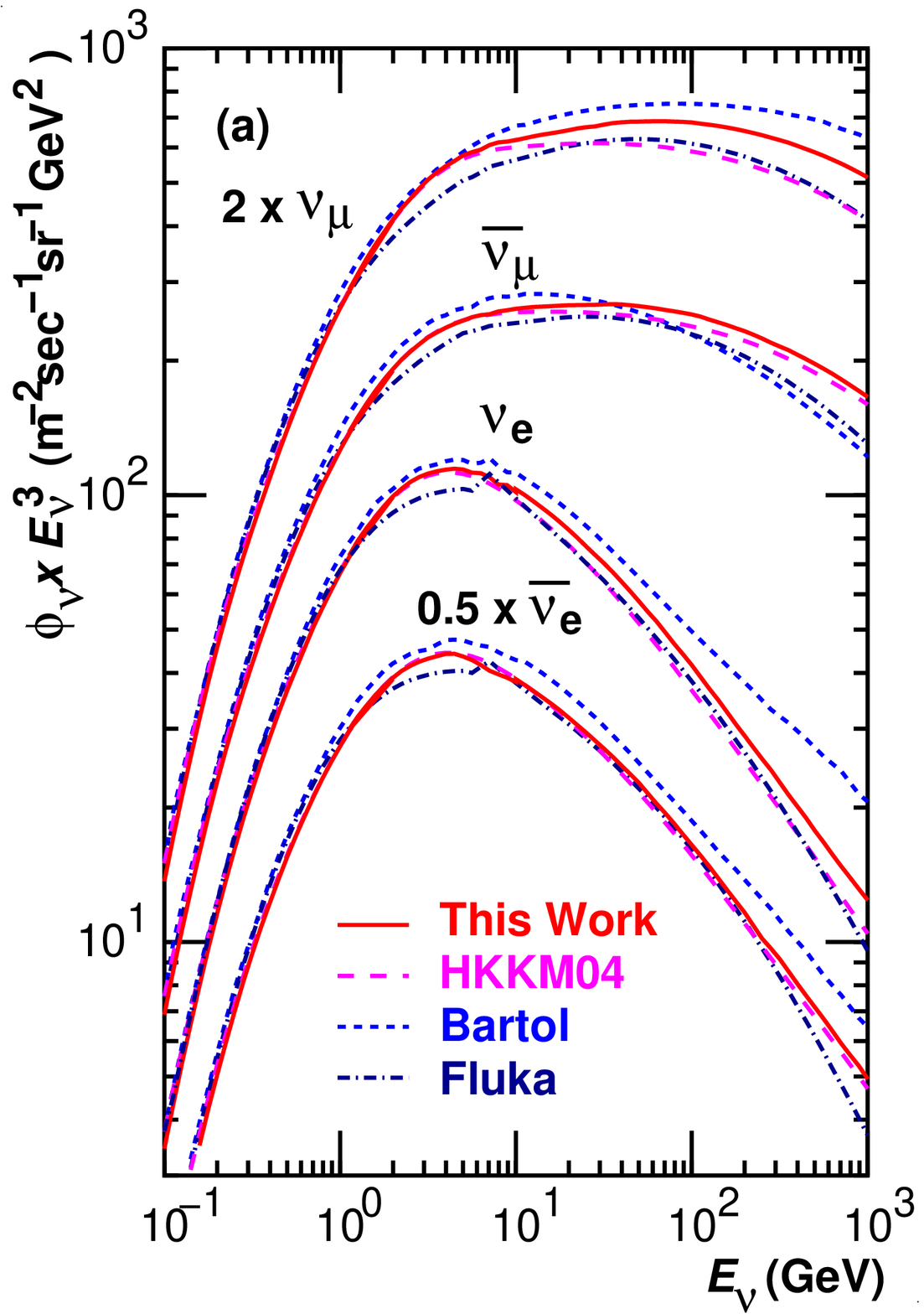}%
\includegraphics[width=5cm]{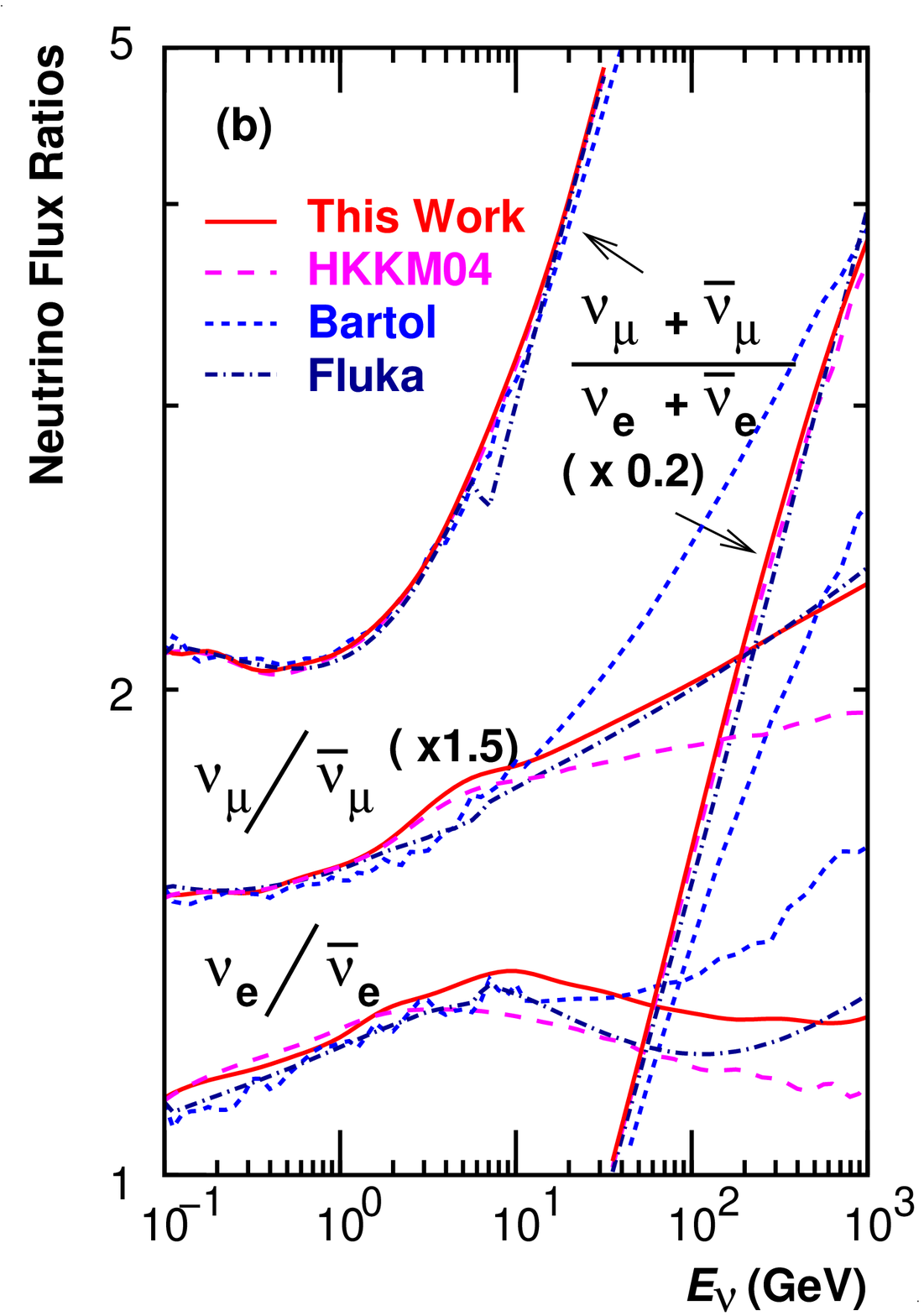}%
}
\caption{\label{fig:other-nflx}
The comparison of all direction average of the atmospheric neutrino fluxes
with other calculations~\cite{barr2004,agrawal1995,battis2002,battis2003}; 
(a) the absolute values of each kind of neutrinos and (b) the ratio of them.
}
\end{figure}

In Fig~\ref{fig:other-nflx}, 
we compared the atmospheric neutrin fluxes calculated with the modified
and original DPMJET-III
with those from other calculations based on the 3-dimensional calculation 
scheme, 
Bartol~\cite{barr2004,agrawal1995} and Fluka ~\cite{battis2002,battis2003}.
The atmospheric neutrino fluxes calculated for Kamioka and averaged 
over all the directions are depicted in panel (a) 
and the ratios are compared in panel (b) up to 1~TeV. 
Although there are sizable differences in the flux values among 
different calculations,
the ratio $(\nu_\mu+\bar\nu_\mu)/(\nu_e+\bar\nu_e)$
is almost identical to each other below 100~GeV among them.

\section{\label{sec:others}Modification of FLUKA'97 and Fritiof 7.02}

In this section, 
we modify the FLUKA'97 and Fritiof 7.02 interaction models,
so that they reproduce the observed atmospheric muon data,
following the procedure we used to modify DPMJET-III in Paper~I.
Then we calculate the resulting atmospheric neutrino flux to study 
the robustness of the modification procedure.
The calculations in this section are carried out using 
the 1-dimensional calculation scheme for computation speed.
The modification is applied to the hadronic interactions above 30~GeV,
to study the muon flux above 10~GeV/c
and neutrino flux above 3~GeV/c.

\begin{figure}[tbh]
\centerline{
\includegraphics[width=8cm]{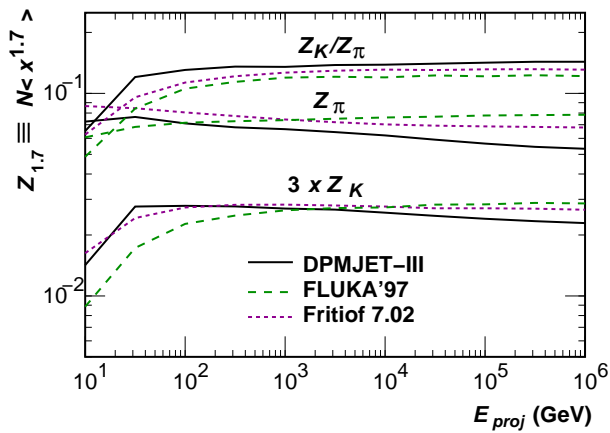}%
}
\caption{\label{fig:z17-orig}
$Z$-factors for  ${\pi^+} +{\pi^-}~(Z_\pi)$ and ${K^+} +{K^-}~(Z_K)$,
and their ratio in each interaction model as the function of projectile 
energy.
Solid lines are for DPMJET-III, 
dashed lines for FLUKA'97,
and dotted lines for Fritiof 7.02.
}
\end{figure}

The secondary spectra of $\pi$ and $K$ productions differ between
interaction models.
The difference is seen in the $Z$-factors defined as
\begin{equation}
Z_i \equiv N_i <x_i^{1.7}>~{\rm and} x \equiv \frac{p_i}{p_{proj}},
\end{equation}
where $N_i$ is the multiplicity and $p_i$ is the momentum
of the $i$ secondary particle, and  $p_{proj}$ is the projectile
momentum.
The $Z$-factors are compared in the sums, 
$Z_{\pi^+} +Z_{\pi^-}$ and $Z_{K^+} +Z_{K^-}$, for 
DPMJET-III, FLUKA'97, and Fritiof 7.02. in Fig.~\ref{fig:z17-orig}.
Note, the contribution of neutral $K$'s to the neutrino flux is 
smaller than that of the charged $K$'s
and $\pi^0$ does not contribute to either muon flux or neutrino flux. 
The neutral $\pi$ and $K$ are not compared in the figure.

\begin{figure}[tbh]
\centerline{
\includegraphics[width=7cm]{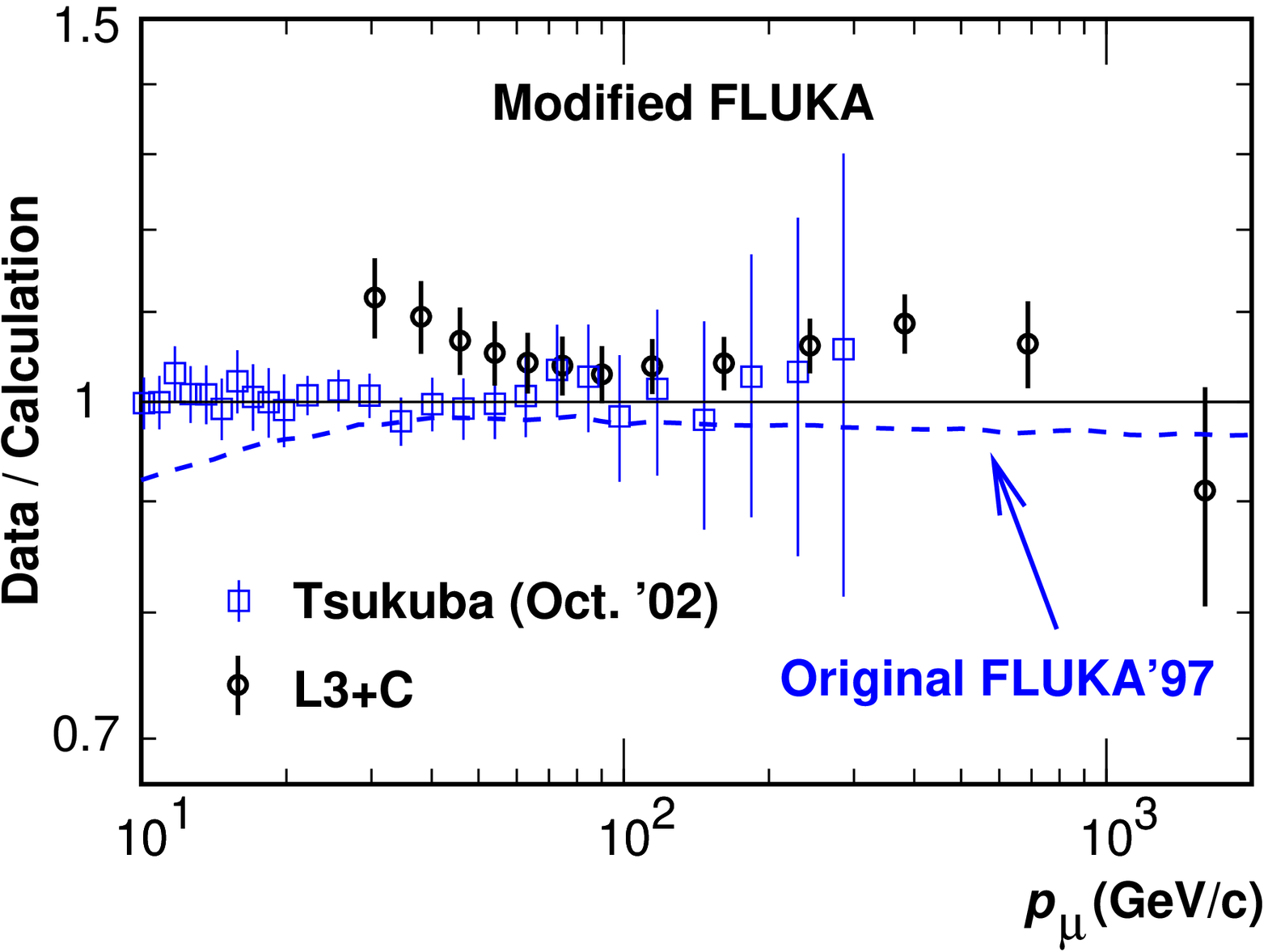}%
\hspace{5mm}
\includegraphics[width=7cm]{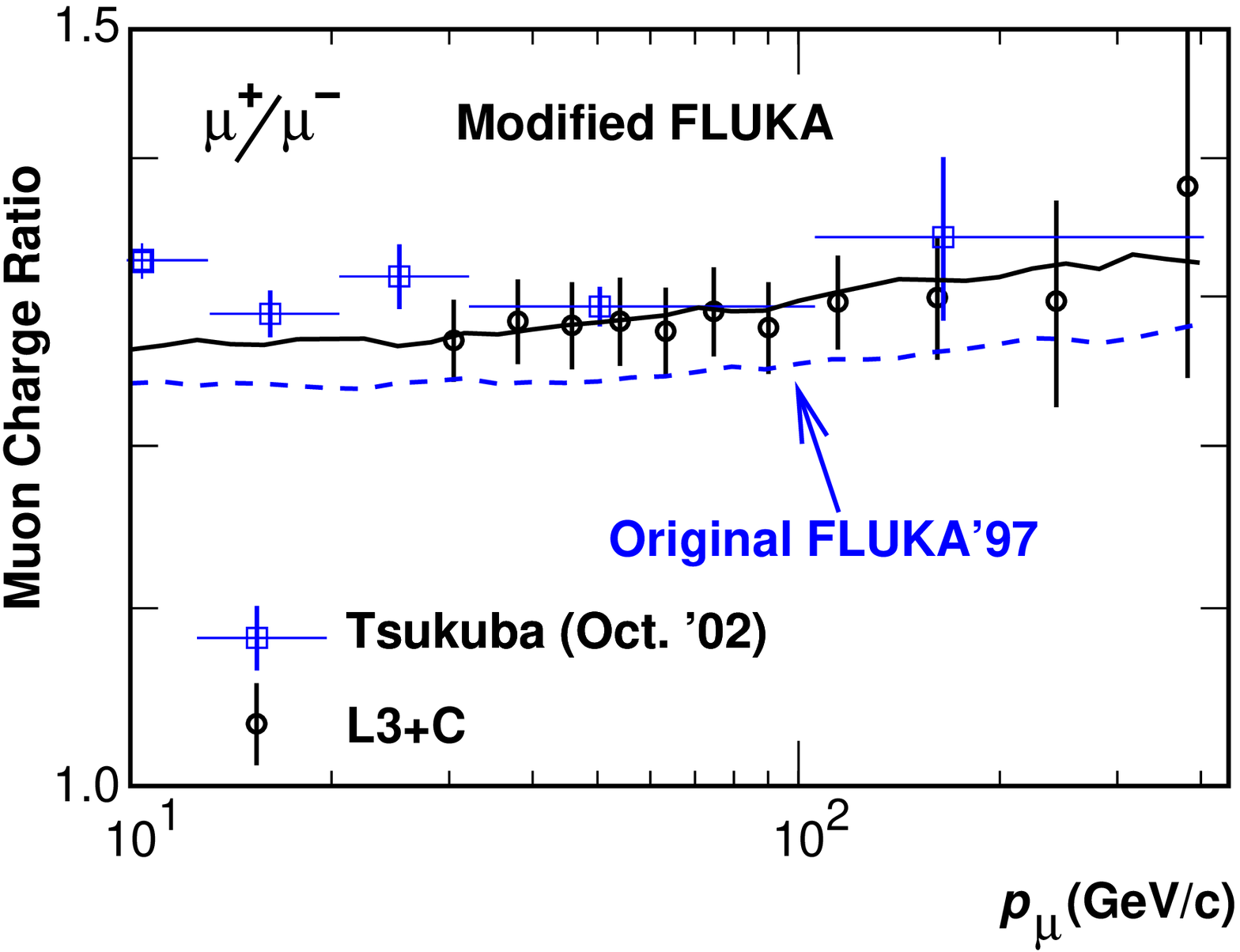}
}
\caption{\label{fig:modified-fluka}
Left: The comparison of muon flux data and the calculations. 
The observed data are shown as the ratios to the calculations
with modified FLUKA.
The dashed line shows the ratio of the calculation with
the original FLUKA'97 to that with modified FLUKA.
Right: Comparison of observed muon charge ratio with the calculations.
The solid line shows the calculation with modified FLUKA
and the dashed line the calculation with original FLUKA'97.
}
\end{figure}

In Fig.~\ref{fig:modified-fluka}, 
we show the comparison of observed muon fluxes and calculations
with the original and modified FLUKA'97 interaction models.
We find the modification clearly improves the agreement 
of the observations and the calculation.
The muon flux calculated with the original FLUKA'97 shows rather 
a better agreement than that of original DPMJET-III above 30~GeV/c
(see also Fig.~\ref{fig:dpm266}).
However the muon flux reproduced by FLUKA'97 becomes
increasingly smaller 
than the observation for momenta below 30~GeV/c.
This feature is also observed in the reconstruction of the muon flux
for balloon altitudes~\cite{Abe:2003ah}, 
and FLUKA'97 was not used in HKKM04.
Hereafter, we refer to this calculation scheme as the ``modified FLUKA''.

\begin{figure}[tbh]
\centerline{
\includegraphics[width=7cm]{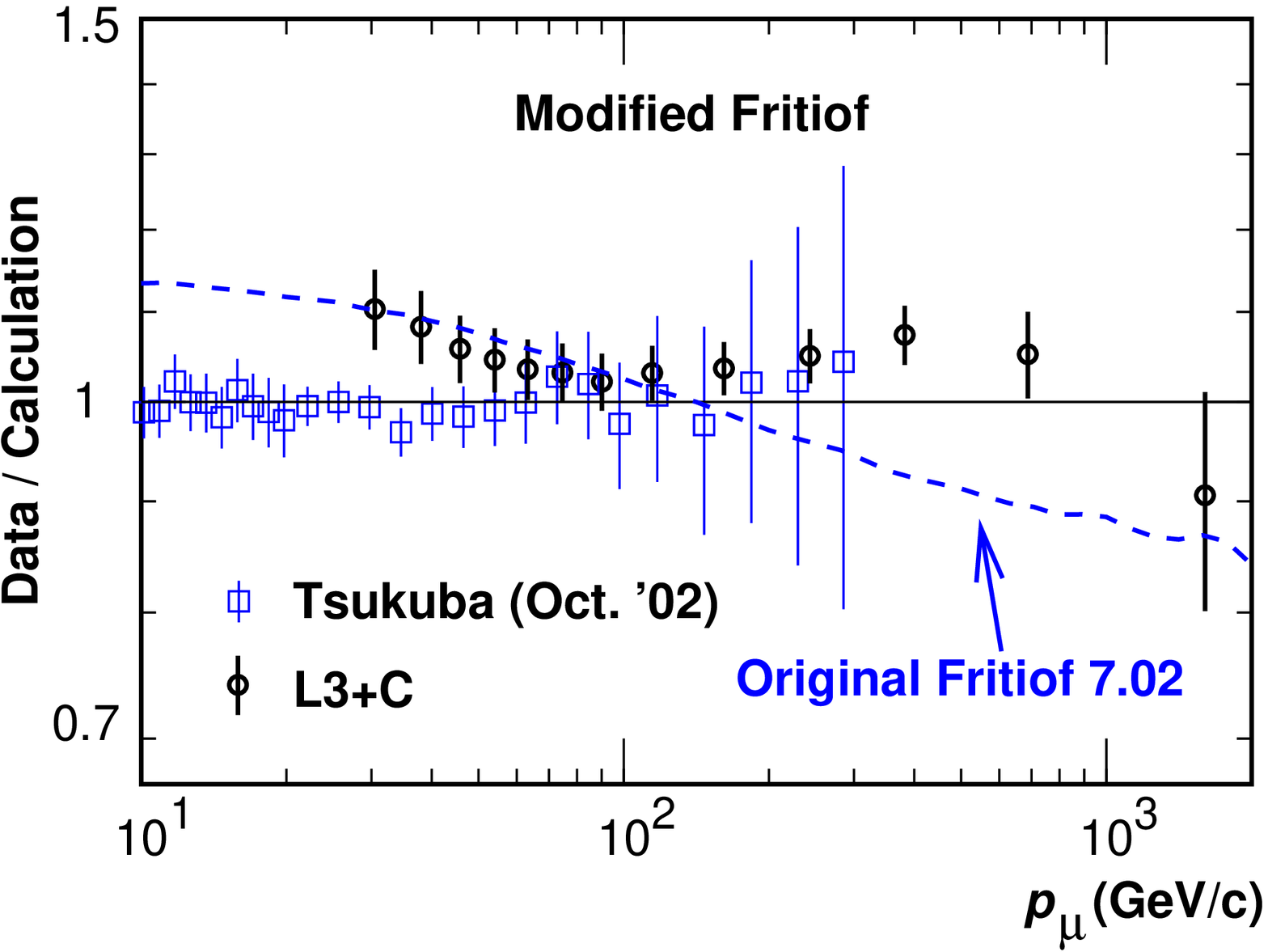}%
\hspace{5mm}
\includegraphics[width=7cm]{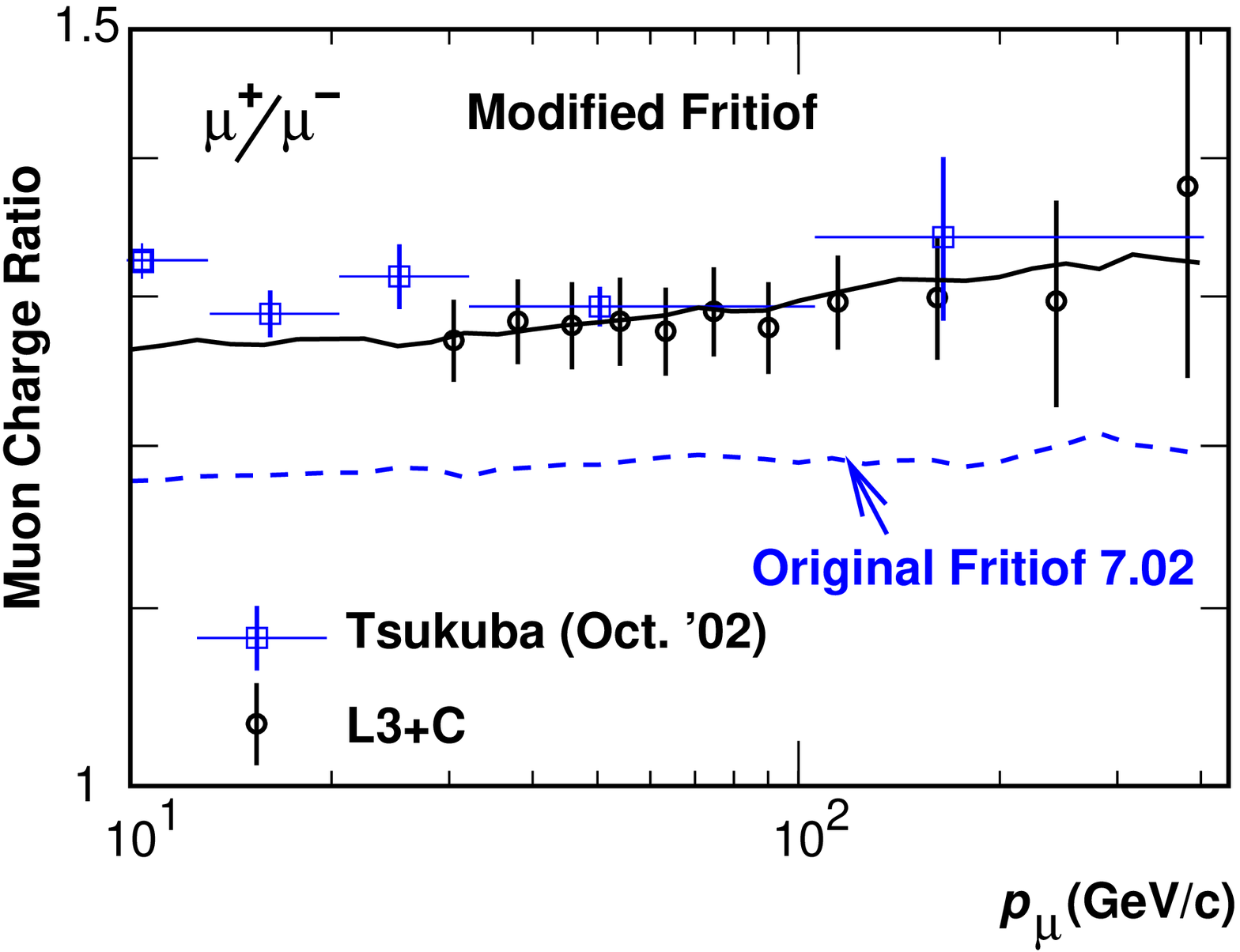}
}
\caption{\label{fig:modified-lund7}
Left: The comparison of muon flux data and the calculations.
The observed data are shown as the ratios to the calculations with 
modified Fritiof.
The dashed line shows the ratio of the calculation with 
the original Fritiof 7.02 to that with modified Fritiof.
Right: Comparison of observed muon charge ratio with the calculations.
The solid line shows the calculation with modified Fritiof
and the dashed line the calculation with original Fritiof 7.02.
}
\end{figure}

In Fig.~\ref{fig:modified-lund7}, 
we show the comparison of observed muon fluxes and the calculations 
with the original and modified Fritiof 7.02 interaction models.
We find here again the modification clearly improves the agreement 
of the observations and the calculation.
The disagreement of the muon flux calculated with the original Fritiof 7.02
is larger than those with the original DPMJET-III or the original FLUKA'97.
We find the modification improved the agreement so that it is almost as good as 
the modified DPMJET-III or modified FLUKA.
Hereafter, we refer to this calculation scheme
as the ``modified Fritiof''.

\begin{figure}[tbh]
\centerline{
\includegraphics[width=7cm]{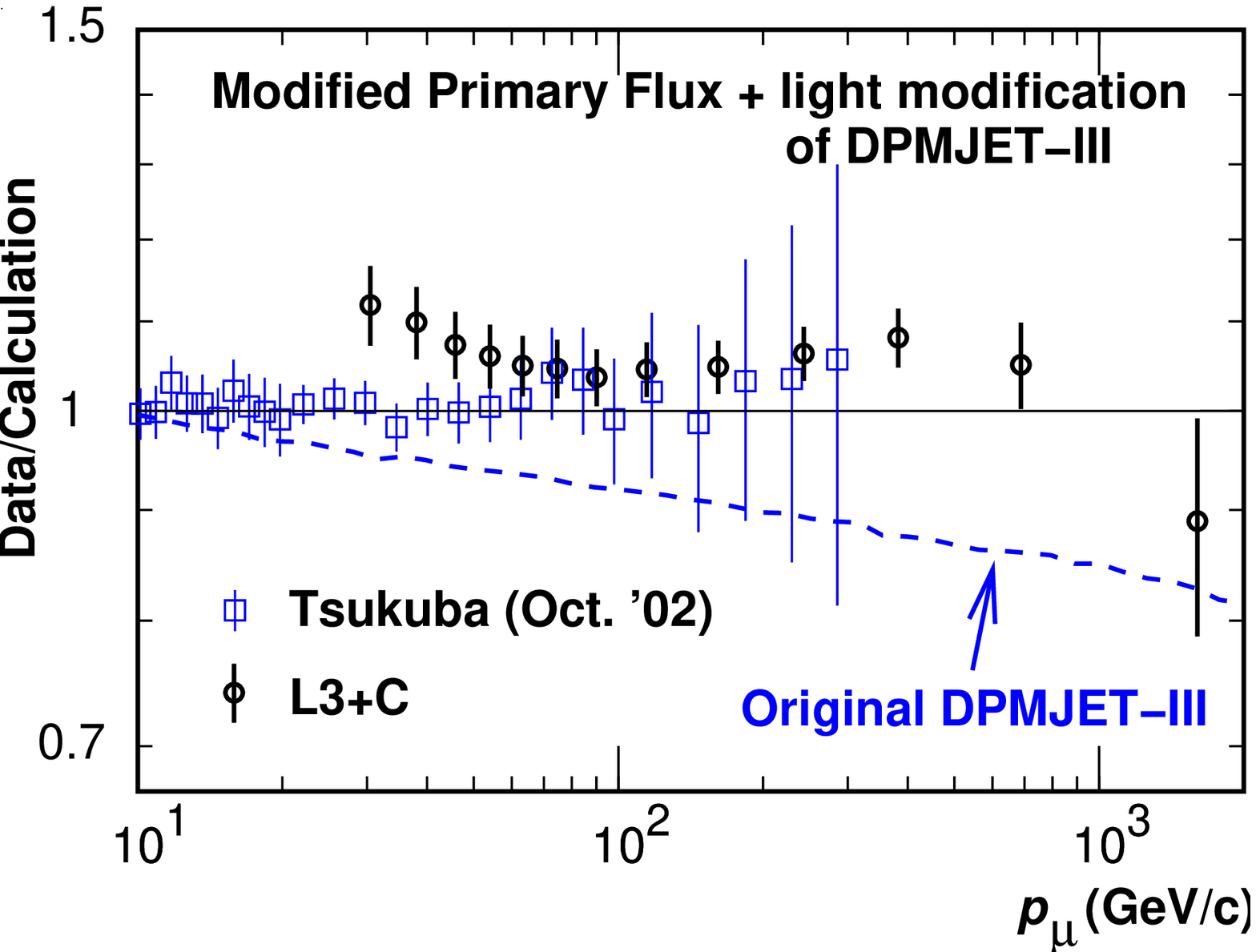}%
\hspace{5mm}
\includegraphics[width=7cm]{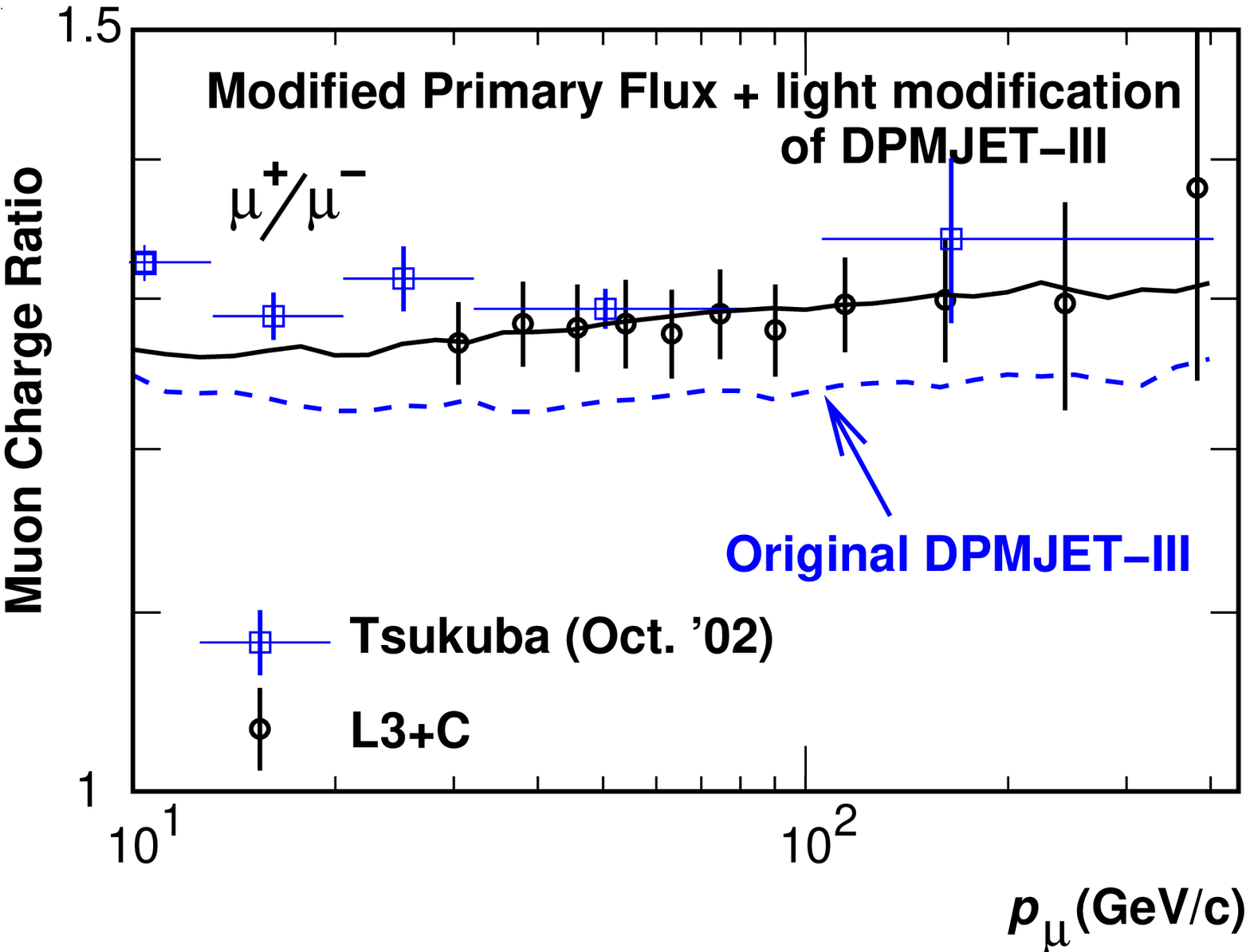}
}
\caption{\label{fig:dpm266}
Left: The comparison of muon flux data and calculations.
The observed data are shown as the ratios to the calculations in
modified primary flux scheme.
The dashed line shows the ratio of calculation in original HKKM04 scheme 
to that in modified primary flux scheme.
Right: The comparison of muon charge ratio between observed data 
and the calculated in modified primary flux scheme.
The solid line shows the calculation with modified primary flux scheme
and the dashed line the calculation in original HKKM04 scheme.
Note, in the modified primary flux scheme, a modification is applied to 
DPMJET-III to reproduce the observed muon charge ratio. For the details, 
see the text. 
}
\end{figure}

With the original DPMJET-III,
we can reproduce the muon flux data in $\mu^+ + \mu^-$ sum
by the modification of the primary flux model
changing of the spectral index from $-$2.71 to $-$2.66 above 
100~GeV~\cite{atic05}.
However, with only the modification of the primary flux model, 
it is difficult to reproduce the observed muon charge ratio.
In terms of $Z$-factor, we find $Z_{\pi^+}/Z_{\pi^-}$ must be 
$\sim 1.5$ at 1~TeV to reproduce the observed muon charge ratio at 100~GeV,
while $Z_{\pi^+}/Z_{\pi^-}$ is $\sim 1.35$ at 1~TeV in the 
original DPMJET-III.
We calculate the atmospheric muon flux with this modified primary
flux model, also applying a light modification for DPMJET-III
to reproduce the observed muon charge ratio, by 
increasing the ratio $Z_{\pi^+}/Z_{\pi^-}$ to $\sim$~1.5.
The calculated results are compared with the observation in 
Fig.~\ref{fig:dpm266}.
We find again the agreement of calculation and observation
are equally as good as the other modified calculations.
Hereafter, we will refer to this calculation scheme 
simply as the ``modified primary flux''.

\begin{figure}[tbh]
\centerline{
\includegraphics[width=8cm]{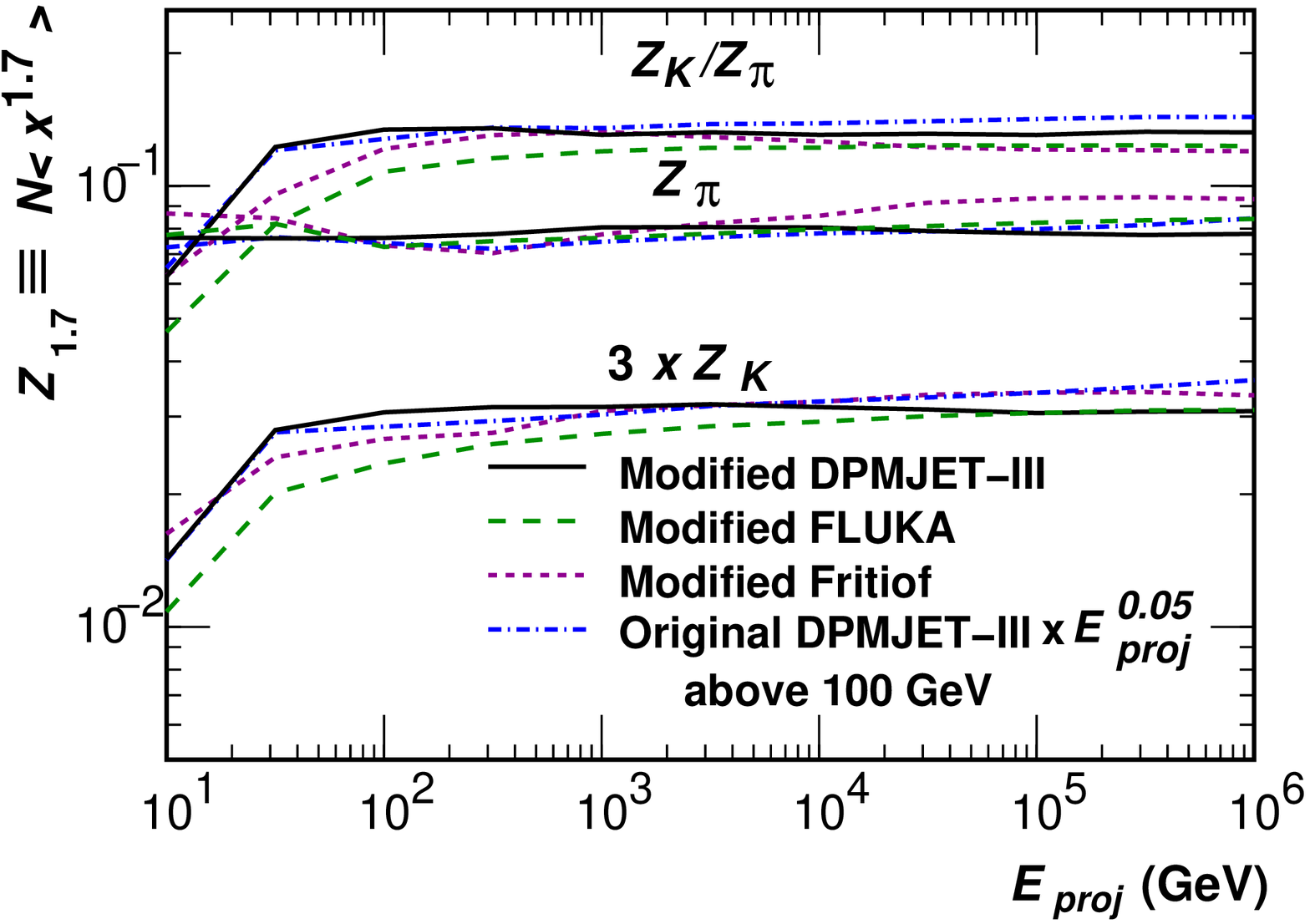}%
}
\caption{\label{fig:z17-tuned}
$Z$-factors for  ${\pi^+} +{\pi^-}~(Z_\pi)$ and ${K^+} +{K^-}~(Z_K)$,
and their ratio in the modified interaction models as the 
function of projectile energy.
Solid lines are for DPMJET-III, 
dashed lines for FLUKA'97,
and dotted lines for Fritiof 7.02.
We also plotted the values of original DPMJET-III 
multiplying $E_{proj}^{0.05}$ above 100~GeV,
to compare the modified primary flux scheme in terms of the $Z$-factor.
}
\end{figure}

In Fig.~\ref{fig:z17-tuned}, we plotted the
$Z_{\pi^+} +Z_{\pi^-}$ and $Z_{K^+} +Z_{K^-}$ for 
the interaction models modified as explained above.
In addition,
we plotted the values of the original DPMJET-III 
multiplying $E_{proj}^{0.05}$ above 100~GeV,
to compare the modified primary flux scheme in terms of the $Z$-factor.
We find all the modified calculations show good agreement in $Z_\pi$
in 30~GeV~$\lesssim E_{proj} \lesssim$~10~TeV,
due to the adjustment of the $\pi$ productions with the muon flux data.
The $Z_K$'s are also closer to each other in the modified calculations.
However, they still show large variations, even in
30~GeV~$\lesssim E_{proj} \lesssim$~10~TeV.
The difference of  $Z_K$'s results from the
original interaction models, and may be considered as 
the reasonable variation of $K$ productions after the modification.

\begin{figure}[tbh] \centerline{
\includegraphics[width=5cm]{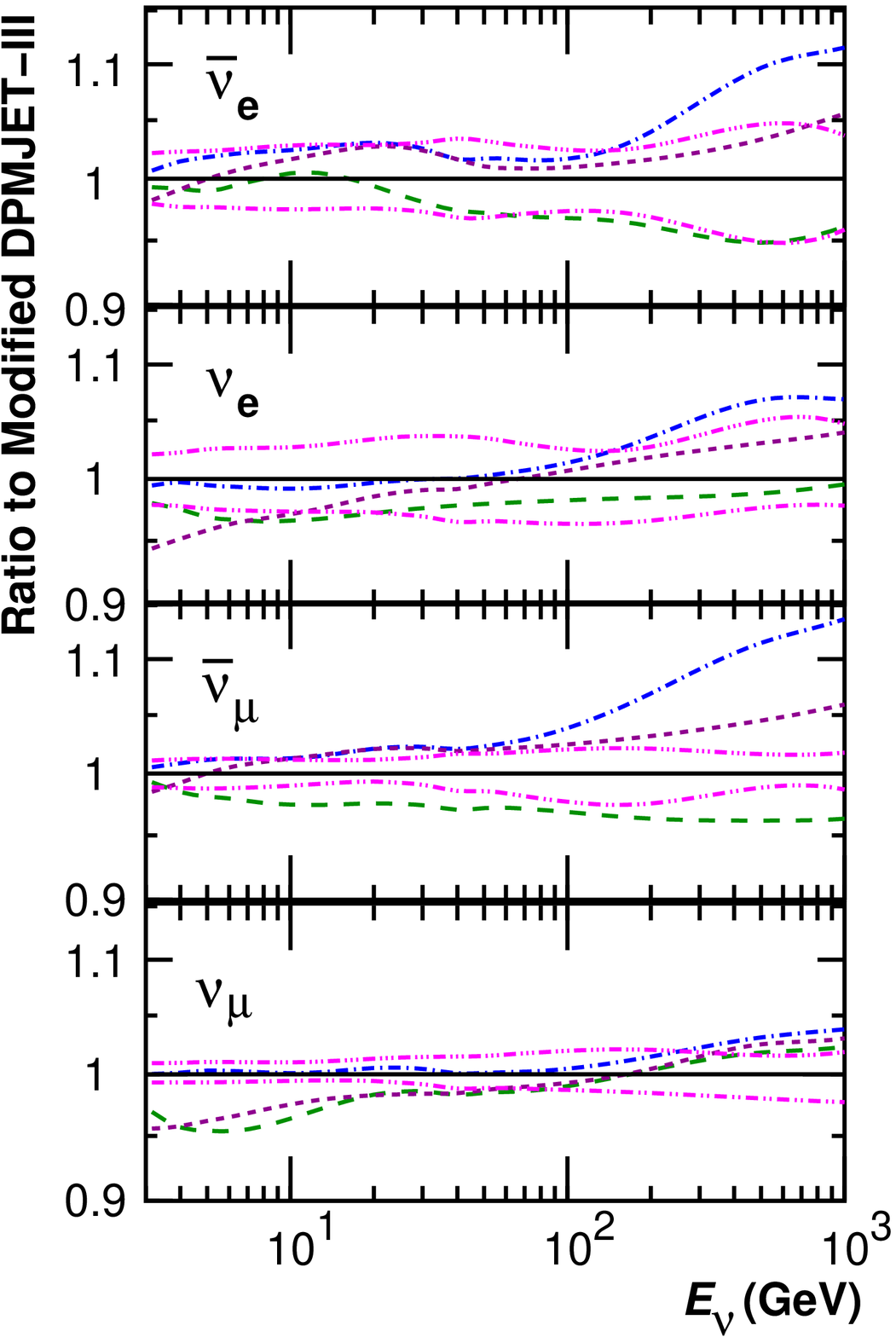}%
\hspace{1cm}
\includegraphics[width=5.18cm]{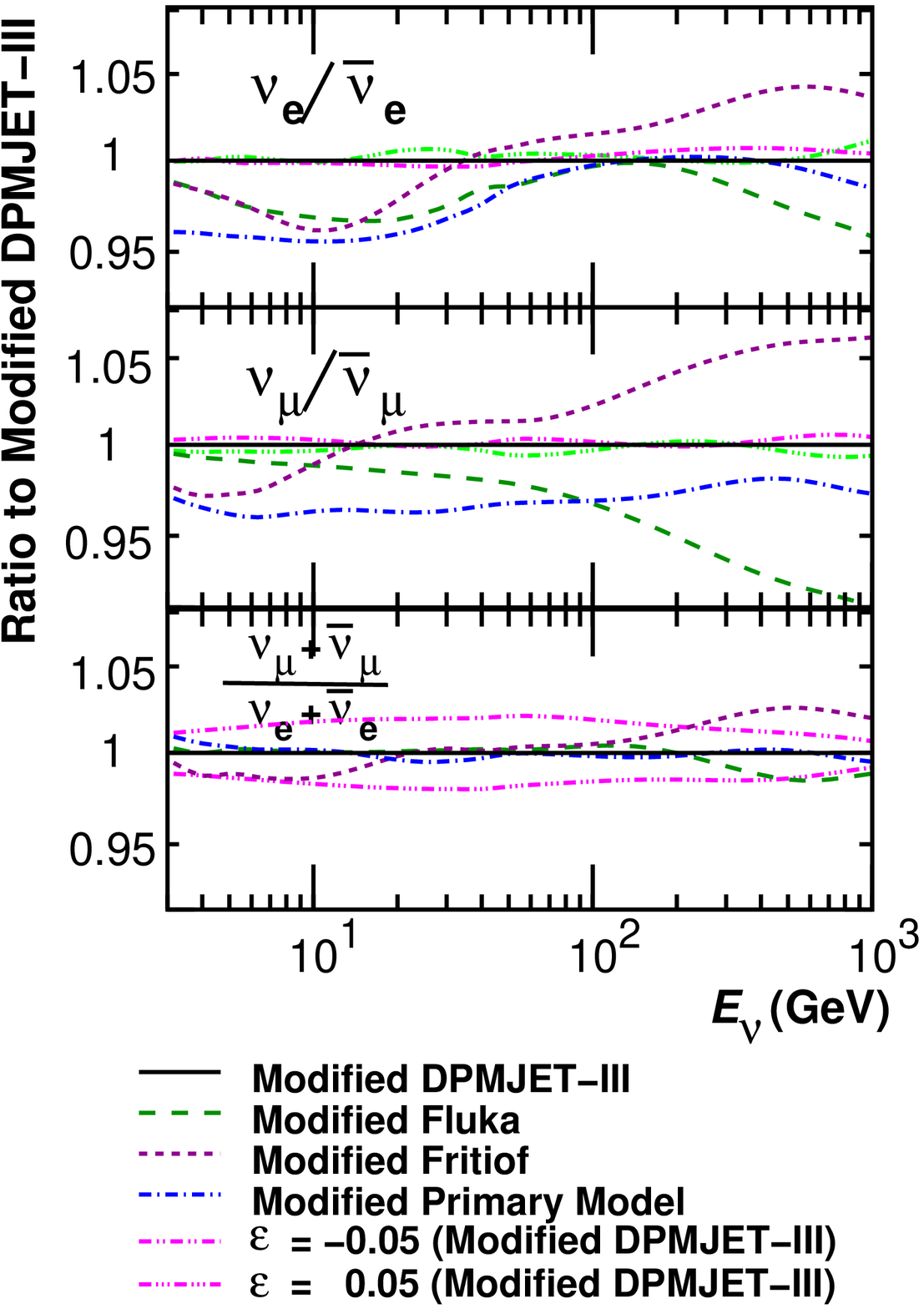}%
}
\caption{\label{fig:comp-modified}
Left panel: comparison of the atmospheric neutrino fluxes 
calculated in the different schemes.
The atmospheric neutrino fluxes are averaged over all directions
and the ratios to that of modified DPMJET-III are shown.
The solid lines are for the modified DPMJET-III,
the dashed lines for the modified FLUKA,
the dotted lines for the modified Fritiof,
and the dash-dot lines for the modified primary flux.
Right panel:  comparison of neutrino ratios, $\nu_\mu/\bar\nu_\mu,
\nu_e/\bar\nu_e$, and $(\nu_\mu+\bar\nu_\mu)/(\nu_e+\bar\nu_e)$.
The neutrino ratios are calculated for all direction averaged flux
in each calculation scheme, 
and the ratio to the modified DPMJET-III shown.
In addition, 
we show the ratios calculated with a slightly different atmospheric model
with $\varepsilon=\pm 0.05$ in Eq.~\ref{eq:scaleheight}.
}
\end{figure}

Now we calculate the atmospheric neutrino flux with the modified calculation
schemes. The calculations are carried out in the 1-dimensional scheme,
and the results are compared with those with modified DPMJET-III
in Fig.~\ref{fig:comp-modified} above 3~GeV.
Note, we also depicted the comparison of the calculations 
with different atmosphere models defined as,
\begin{equation}
\label{eq:scaleheight}
\rho_{us,\varepsilon}(h) = \frac 1 {1+\varepsilon} \cdot \rho_{us}(\frac h {1+\varepsilon})~,
\end{equation}
where $\rho_{us}(h)$ is the density profile of 
US-standard '76 \cite{us-standard}.
This is the same modification of the US-standard '76, which we used
in Sec.~IV of Paper~I to study the effect of the atmospheric density 
profile on the lepton fluxes.
We take the parameter $\varepsilon$ to be $\pm$~0.05 as in that study, 
and use the modified DPMJET-III for the interaction model.

In the left panel of  Fig.~\ref{fig:comp-modified}, we find
the atmospheric neutrino fluxes calculated with the 
modified calculations agree within $\pm$~5~\% below 100~GeV.
The agreement is considered to be due to the large contribution 
of $\pi$'s in this energy region. 
Note, the different atmospheric models result in 
almost the largest difference for $\nu_e$ and $\bar \nu_e$,
and the difference for $\nu_\mu$ and $\bar \nu_\mu$
is smaller than other modified calculations.
This is a result of differences in $\pi - \mu$ successive decay and the 
muon propagation in the atmosphere.

Above 100~GeV, where the contribution of $K$'s becomes large,
we find a large variation of neutrino fluxes for all kinds of 
neutrino except for the $\nu_\mu$ flux.
In our modification scheme, the $K^+$ productions are modified 
at the same rate as the $\pi^+$ productions.
Therefore, the variation of $\nu_\mu$ flux is relatively 
small even above 100~GeV.
Note,
the modified primary flux scheme produces the largest 
neutrino fluxes among all the modified calculations and 
for all kinds of neutrino except for the $\nu_\mu$ above 100~GeV.
This is due to the increase of $K^-$ productions by 
the increase of primary cosmic ray in this model.
In our modification scheme, the $K^-$ productions are not altered.

In the right panel of Fig.~\ref{fig:comp-modified}, 
we show the neutrino 
flux ratios, $\nu_\mu/\bar\nu_\mu,\ \nu_e/\bar\nu_e$, 
and $(\nu_\mu+\bar\nu_\mu)/(\nu_e+\bar\nu_e)$, 
averaged over all directions.
We find the largest difference in the $\nu_\mu/\bar\nu_\mu$ ratio,
and it could be explained by the differences in $\nu_\mu/$
and $\bar\nu_\mu$ seen in the left panel, 
and by the $K$-productions at higher energies.
The difference of the ratios in the 3$\sim$100~GeV range
is much less than $\pm$ 5 \% for all ratios,
where the contribution of $\pi$'s is important. 
Especially, the difference in $(\nu_\mu+\bar\nu_\mu)/(\nu_e+\bar\nu_e)$ ratio
is small at all the energies shown here ($\lesssim$ 2 \%).
The largest variation in the $(\nu_\mu+\bar\nu_\mu)/(\nu_e+\bar\nu_e)$ ratio
is found with the change of atmospheric model.
This is the direct consequence of the large variation of 
$\nu_e$ and $\bar \nu_e$ and small variation $\nu_\mu$ and $\bar \nu_\mu$
resulting from the change in the atmospheric model.

\section{\label{sec:systematic} Uncertainty in the Atmospheric neutrino Flux calculation}
%modulation is not considered here

\subsection{Uncertainty for flux value of the neutrinos.}

Here, we estimate the total uncertainty or the total possible errors 
in the calculation of atmospheric neutrino flux.
It may be expressed as
\begin{equation}
\label{eq:total-error}
\delta_{\rm tot}^2  
= \delta_{\pi}^2 
+ \delta_{K}^2 
+ \delta_{\sigma}^2 
+ \delta_{\rm air}^2  
+ (\delta_{\rm scheme}^2  
+ \delta_{\rm stat}^2
+ \cdots), 
\end{equation}
where $\delta_{\pi}$  is the uncertainty due to the uncertainty of
$\pi$ production in the hadronic interaction model,
$\delta_{K}$ is due to the $K$ production,
$\delta_{\sigma}$ due to the hadronic interaction cross sections,
$\delta_{\rm air}$ due to the atmospheric density profile,
$\delta_{\rm scheme}$ due to the calculation scheme including any
bugs in the code, and 
$\delta_{\rm stat}$ due to statistical errors.
The solar modulation of the cosmic rays and mountains above the
neutrino detector cause sizable effects on the atmospheric 
neutrino flux. However, they are not a true uncertainty and 
they are included in our calculation correctly.

The statistical error in the Monte Carlo study 
is smaller than 1\% below 3~TeV and 3\% below 10~TeV for 
$\nu_\mu$ and $\bar\nu_\mu$. 
For $\nu_e$ and $\bar\nu_\mu$, it is a little worse and smaller 
than 1\% below 10~GeV, 3\% below 3~TeV, and around 10\% at 10 TeV.
However, the statistical error is much smaller than
those from other uncertainty sources.
The largest error due to the calculation scheme is the finite 
size effect of the virtual detector, which we studied in detail 
in Sec.~\ref{sec:calculation}.
With the procedure proposed there, the remaining error would be
much smaller than 1\%.
We do not discuss $\delta_{\rm scheme}$ and $\delta_{\rm stat}$ 
in the following.

\begin{figure}[tbh]
\includegraphics[width=7cm]{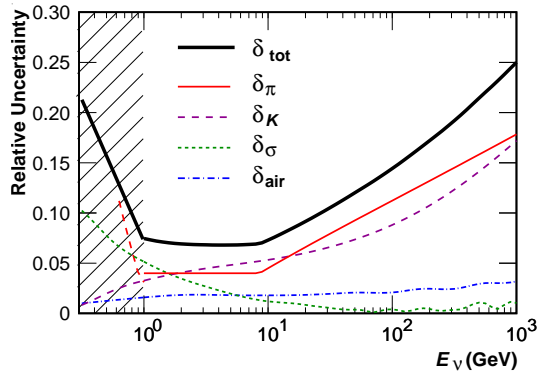}%
\caption{\label{fig:errnu-tot}
The uncertainty of each error source for atmospheric neutrino 
flux and their sum with Eq.~\ref{eq:total-error}.
Note, Eq.~\ref{eq:mu-nu} loses its validity
in the shaded region. The total error for $\lesssim$~1 GeV
is estimated differently from 
Eq.~\ref{eq:total-error}, as stated in the text.
Note the statistical and systematic error are not shown in the figure.
}
\end{figure}

In Paper~I, we have proven that the 
error of $\pi$ productions in the hadronic interaction model 
affects the atmospheric muon and neutrino fluxes produced by the 
$\pi$ decay at the same rate, namely
\begin{equation}
\label{eq:mu-nu}
\frac {\Delta\phi_\mu}{\phi_\mu} \simeq 
\frac{\Delta\phi_{\nu_\mu}}{\phi_{\nu_\mu} }
\simeq \frac{\Delta\phi_{\nu_e}}{\phi_{\nu_e}}
\end{equation}
above 1~GeV. 
We may estimate $\delta_\pi$ from the comparison of the 
observation and calculation of the atmospheric muon flux.
Since the atmospheric muon flux below 1~TeV comes almost only from the 
$\pi$ decay,
we use the sum of the experimental error and the residual of 
the reconstruction 
as the $\Delta\phi_\mu$ in Eq.~\ref{eq:mu-nu} (see Fig.~15 of Paper~I).
Then we replace $\delta_{\pi}$ in Eq.~\ref{eq:total-error} with 
$(\Delta\phi_\mu/\phi_\mu)\phi_\nu$, where $\phi_\nu$ is the sum of 
$\pi$ and $K$ contributions for a conservative estimation.
The estimated uncertainty is depicted by the solid line above 1~GeV 
in Fig.~\ref{fig:errnu-tot}.

For the $\delta_{K}$, we used the modified calculation schemes studied 
in  Sec.~\ref{sec:others}.
We assumed the maximum neutrino flux difference from the modified DPMJET-III
among them as $\delta_{K}$.
The maximum difference for all kinds of 
neutrino for vertical direction is depicted by the dashed line
in Fig.~\ref{fig:errnu-tot}, since that variation is the largest of all
zenith angles.
Each difference is a little larger, but similar to that shown 
in the left panel of Fig.~\ref{fig:comp-modified}.
Note, the maximum difference from the modified DPMJET-III 
is seen in the modified primary flux model in most of the cases.

For $\delta_{\sigma}$,
we assumed the difference $|\Delta \phi_\mu - \Delta \phi_\nu |$ 
in the Fig.~10 of Paper~I.
Since the uncertainty of the interaction cross section works
with opposing effects for atmospheric muons and neutrinos,
the error of the interaction cross section introduces an error
in the calibration of interaction model with the atmospheric 
muon flux data.
On the other hand, 
as we use the observed atmospheric density profile,
the calibration is not affected by the error of the atmospheric model.
We use $\Delta\phi_\nu$ only in Fig.~9 of Paper~I as 
the $\delta_{\rm air}$.
All these uncertainties, $\delta_{\pi}$($\delta_{\mu}$), 
$\delta\phi_{K}$, $\delta\phi_{\sigma}$, $\delta\phi_{\rm air}$,
and $\delta_{\rm tot}$, 
are summarized in Fig.~\ref{fig:errnu-tot}.
Note, the estimations are conservative, and the maximum uncertainty
is shown for all kind of neutrinos and zenith angles.

\begin{figure}[tbh]
\centerline{
\includegraphics[width=6.5cm]{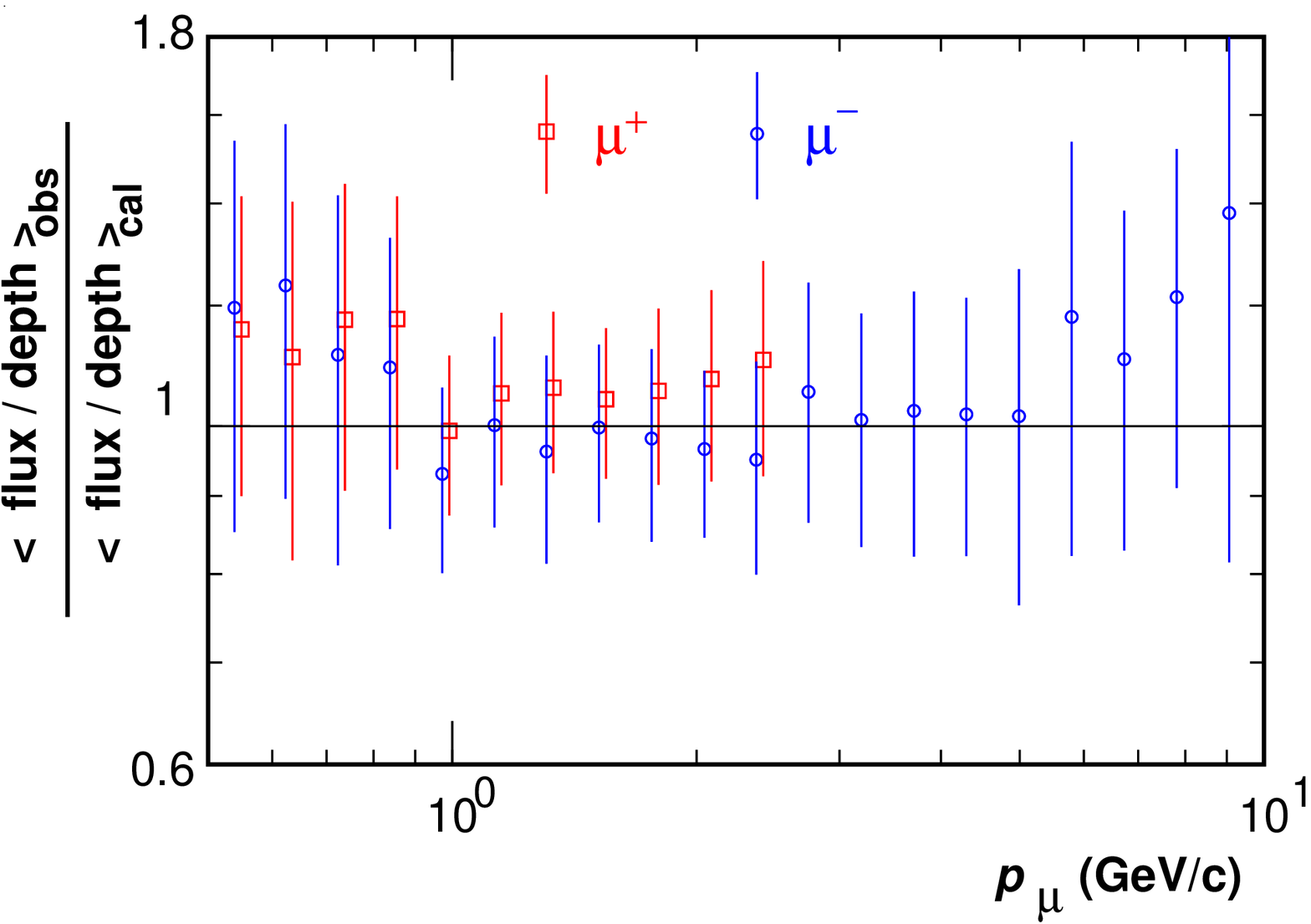}%
~
\includegraphics[width=6.05cm]{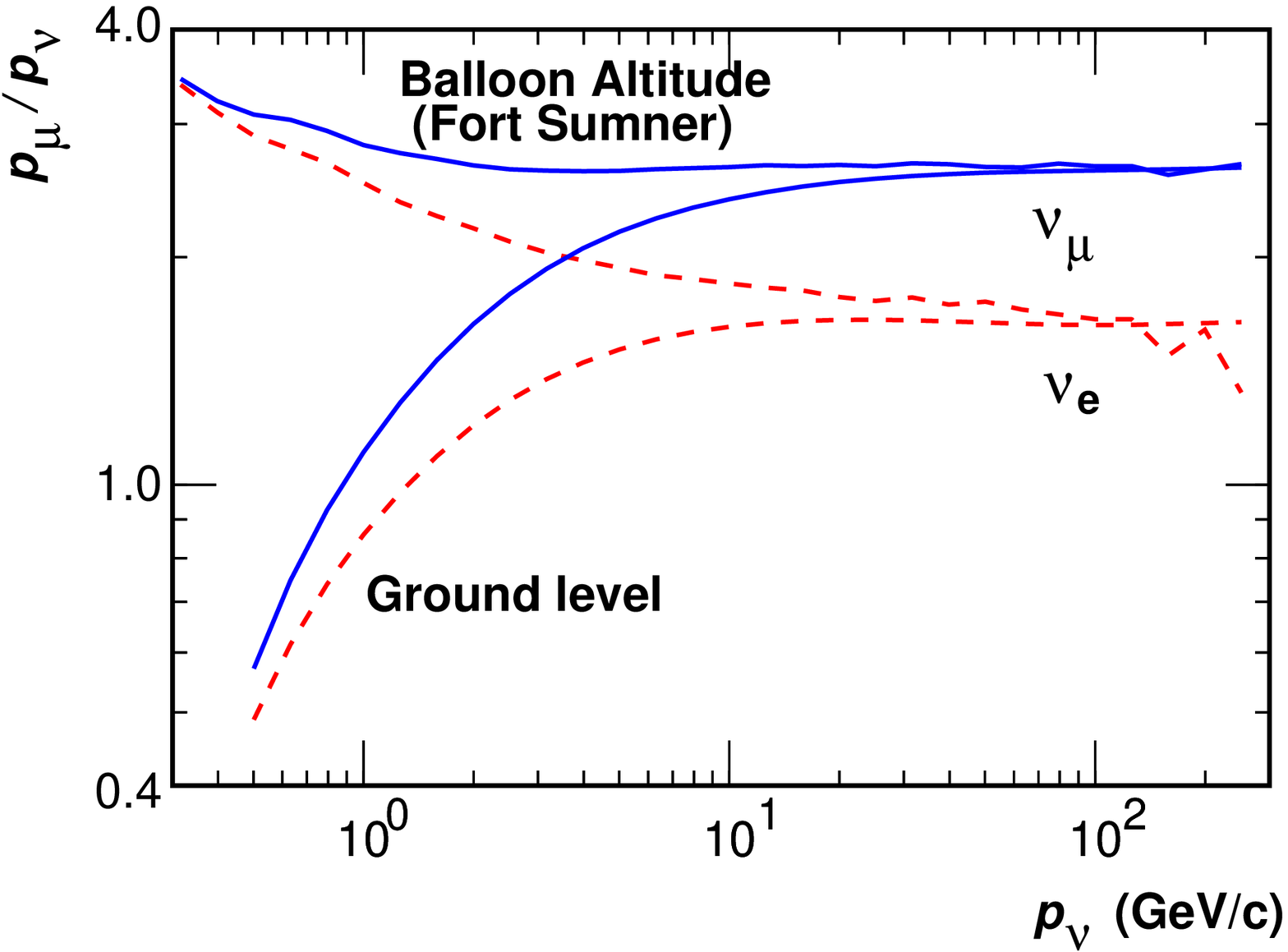}%
}
\caption{\label{fig:fort-mu}
The muon fluxes at balloon altitudes at Fort Sumner (Sept.\ 2001).
Left: comparison calculated and observed muon fluxes.
The dashed lines show $\pm$~10\% deviation from the calculated values.
Right: $p_\mu/p_\nu$ ratio for the same momentum of parent $\pi$'s in average.
The dashed lines are the same quantities but for the muons at the ground level.
}
\end{figure}

We note, Eq.~\ref{eq:mu-nu} is valid only for $\gtrsim$~1 GeV.
We have to estimate $\delta_\pi$ without using the atmospheric
muon flux data at ground level.
In Fig.~\ref{fig:fort-mu}, we show the study of the muon flux
at balloon altitudes at Fort Sumner~\cite{Abe:2003cd}.
The modified DPMJET-III reproduces the muon flux within $\pm$~10\% 
at $\sim$~1~GeV/c, and 
$p_\mu/p_\nu$ ratio for the same momentum of parent $\pi$'s
remains $\sim$3 even at the lower momenta,
due to the small energy loss of muons at balloon altitudes.
However, the distance of the production and observation places
are longer than the muons observed at ground level.
The muon decay in this distance make
Eq.~\ref{eq:mu-nu} less accurate for $\lesssim$~1~GeV.
We conservatively estimate 20\% errors for pion productions 
responsible to the atmospheric neutrino at $\sim$0.3~GeV.

Note, the uncertainty studied above is for all the kind of neutrinos,
and for all zenith angles.
Limiting the kind of neutrino and the zenith angle, we may get a smaller
estimation for the uncertainty.
Especially, the uncertainties in the ratio of the different kind of neutrinos
and the zenith angle dependence show smaller uncertainties.
They are important in the study of the
neutrino oscillations using the atmospheric neutrinos,
and are studied in the following.

\subsection{Uncertainty for the flux ratio among different neutrinos}

We have studied the variation of the ratios,
$\nu_\mu/\bar\nu_\mu, \nu_e/\bar\nu_e$, and 
$(\nu_\mu+\bar\nu_\mu)/(\nu_e+\bar\nu_e)$,
among the modified calculations for all direction average 
in Sec.~\ref{sec:others},
and shown in the right panel Fig.~\ref{fig:comp-modified}.
We find the variation of 
the $\nu_\mu/\bar\nu_\mu$ and $\nu_e/\bar\nu_e$ 
is within $\pm$~5\% below 100~GeV, where $\pi$'s are still the 
major source of atmospheric neutrinos.
However, the variation increases above 100~GeV, due to the
difference of the $K$ productions in the different models.

On the other hand,
the ratio $(\nu_\mu+\bar\nu_\mu)/(\nu_e+\bar\nu_e)$ is very stable 
among different calculations especially below 100~GeV ($\lesssim$~2\%).
Note, however, the variation of atmospheric model gives almost the largest 
variation to this ratio in this energy region.
This is explained by the $\pi-\mu$ successive decay process:
\begin{equation}
\begin{array}{c l}
\pi^\pm \rightarrow& \mu^\pm + \nu_\mu(\bar\nu_\mu) \\
                   & \rightarrow e^\pm + \nu_e(\bar\nu_e) + \bar\nu_\mu(\nu_\mu) \\
\end{array}
\end{equation}
The decay and energy loss of muons are affected by the atmospheric density profile.
When most of the muons decay in the atmosphere, 
the ratio  $(\nu_\mu+\bar\nu_\mu)/(\nu_e+\bar\nu_e)$ becomes small.
When muons lose energy at a larger rate, 
the products of muon decay has lower energies, then with the steep energy 
spectrum of $\pi$'s at decay, the ratio becomes larger.
The interaction model also 
causes an uncertainty in 
the ratio $(\nu_\mu+\bar\nu_\mu)/(\nu_e+\bar\nu_e)$ through the 
$\pi$'s spectrum at the decay. However, this is not a large effect as is
seen from the right panel of Fig.~\ref{fig:comp-modified}.

We estimate the error of the ratio $(\nu_\mu+\bar\nu_\mu)/(\nu_e+\bar\nu_e)$ 
is $\sim~\pm$2\% below 100~GeV, since 
our atmospheric density profile is a good approximation
for the one year average of the more realistic model.

\subsection{Uncertainty for the vertical$/$horizontal ratio}

In the determination of $\delta m^2_{23}\equiv m^2_{\nu_3} -m^2_{\nu_2}$
with atmospheric neutrino flux,
the zenith angle dependence of the atmospheric neutrino flux is important.
For the study of the zenith angle dependence,
we calculate the quantity defined as:
\begin{equation}
I_{i}^{(n)} (\cos\theta)
= \int_{E1}^{E2} \phi_i(\cos\theta, E_\nu)  E_\nu^n  dE_\nu~,
\label{eq:stopnu}
\end{equation}
for $i=\nu_\mu,\bar\nu_\mu$, following HKKM04.
The quantity $I^{(2)}_{\nu_\mu}$ and $I^{(2)}_{\bar\nu_\mu}$ are 
roughly proportional to the rate of the 
muon events induced by $\nu_\mu$ and $\bar\nu_\mu$
in the energy range from $E_1$ to $E_2$.  
The neutrino cross section is roughly proportional to the neutrino energy 
and the muon range, therefore, the target volume is proportional to 
the muon energy, as far as the energy loss due to the pair creations is smaller
than the ionization energy loss ($\sim 500$~GeV).
With $E_1=$10~GeV and $E_2$=1~TeV for the integration,
we calculated $I^{(2)}_V$ as the average of $I_{\nu_\mu}^{(2)} + I_{\bar\nu_\mu}^{(2)}$
over $\cos\theta_{z}=0.9-1.0$,
and $I^{(2)}_H$ as the average over $\cos\theta_{z}=0.0-0.1$.
These quantities are calculated in all the modified calculation schemes
studied in Sec.~\ref{sec:others} and in the original ones.
They are summarized in Table~\ref{tab:hov}.

The stability of the zenith angle dependence of atmospheric neutrino flux
may be studied with the stability of the $I^{(2)}_H/I^{(2)}_V$ ratio.
We find $I^{(2)}_H$ and $I^{(2)}_V$ calculated with the modified DPMJET-III
show increases from those calculated in HKKM04 by about 10\%.
Therefore, the expectation value for the neutrino induced muon event
will be also increased by 10\%.
However the $I^{(2)}_H/I^{(2)}_V$ ratio is almost the same as the HKKM04 calculation.
The situations are similar for other interaction models, 
and there are $\sim$~3\% variations in the $I^{(2)}_H/I^{(2)}_V$ ratios
among the different calculations.
Note, the different atmospheric density profile also gives 
a $\sim~\pm$0.8\% variation.
Those variations remain as the uncertainty for the $I^{(2)}_H/I^{(2)}_V$ ratio,
or the uncertainty of the zenith angle dependence of the atmospheric 
neutrino flux.

\begin{table}[htb]
\caption{Quantities calculated by Eq.~\ref{eq:stopnu} 
with $E_1=10$ GeV, $E_2=1$~TeV, and n=2.  
$I_V^{(2)}$ is the sum of $i=\nu_\mu$ and $\bar\nu_\mu$
for the vertical directions ($\cos\theta_{z}=0.9-1.$),
and $I_H^{(2)}$ for the horizontal directions 
($\cos\theta_{z}=0.-0.1$),
where $\theta_z$ is the zenith angle.
Note, the original DPMJET-III is used in the HKKM04 calculation.
}
\vspace*{2truemm}
\begin{tabular}{|c| c| c| c |}% \hline
\toprule
Calculation 
&~$I^{(2)}_V({\rm m^{-2}sec^{-1}GeV^2})$~
&~$I^{(2)}_H({\rm m^{-2}sec^{-1}GeV^2})$~
&\hspace{10mm}$I^{(2)}_H/I^{(2)}_V$\hspace{10mm}~\\
\toprule
\botrule
%\hrule
HKKM04
& 1634.
& 3775.
& 2.310
\\
HKKM04~($\varepsilon=+0.05$)
& 1645.
& 3881.
& 2.285
\\
HKKM04~($\varepsilon=-0.05$)
& 1577.
& 3782.
& 2.322
\\
modified~DPMJET-III
& 1798.
& 4186.
& 2.328
\\
modified~Flux~model
& 1835.
& 4295.
& 2.340
\\
FLUKA'97
& 1689.
& 4068.
& 2.409
\\
modified~FLUKA
& 1725.
& 4151.
& 2.407
\\
Fritiof~7.02
& 1812.
& 4133. 
& 2.281
\\
modified~Fritiof
& 1826.
& 4183.
& 2.290
\\
\botrule
\end{tabular}
\label{tab:hov} 
\end{table}

\section{\label{sec:summary}Summary}

We have revised the calculation of atmospheric neutrino flux in 
HKKM04 with the ``modified DPMJET-III'' constructed in Paper~I.
Before the calculation, we have studied the error caused by the use of 
the ``virtual detectors''
generally far larger than the actual neutrino detectors.
Such a virtual detector introduces an error, 
since it averages the neutrino flux over positions where
the geomagnetic conditions are different from the position of
target detector.
At the low energy end of our calculation (0.1~GeV), 
the error reaches around 5~\% with the virtual detector
used in HKKM04 (a circle with radius $\sim$~1000~km),
but it could be corrected comparing the average flux values
of virtual detectors with different sizes.
With this correction the error could be reduced to $\lesssim$~1~\%
which is much smaller than the uncertainties in other 
components for the calculation of atmospheric neutrino flux.
In the calculation, we updated the geomagnetic field model 
from epoch 2000 to 2005.

Next, we studied the robustness of the modification procedure
explained in Paper~I.
We modified the interaction models of FLUKA'97 and Fritiof 7.02,
so that they reproduce the atmospheric muon flux data (i.e., the 
same as the 
modified DPMJET-III),
and calculated the atmospheric neutrino flux in the 1-dimensional scheme for 
the energies above 3~GeV.
The neutrino fluxes calculated with those modified interaction models 
agree within $\pm$5\% 
with the flux calculated with the modified DPMJET-III below 100~GeV.
However, the variation increases to around $\pm$15\% at 1~TeV.
This is considered to be due to the differences of the $K$ productions in the
modified interaction model.
As the $K$ productions in the original interaction
models are different, there remain the differences of the
$K$ production in the modified interaction models.
With this variation of the neutrino fluxes with the modified interaction
models, we estimated the uncertainty of the atmospheric neutrino flux
due to the uncertainty of the $K$ production in the interaction model,
when they are modified to reproduce the observed muon flux correctly.

We summarized the uncertainties in the calculation of atmospheric 
neutrino flux including that of $K$ production in the interaction model.
The relation 
$\Delta\phi_\mu/\phi_\mu \simeq \Delta\phi_{\nu_\mu}/\phi_{\nu_\mu} 
\simeq \Delta\phi_{\nu_e}/\phi_{\nu_e}$,
for the atmospheric lepton fluxes from the $\pi$ decay
derived in Paper~I is useful in this study.
As we reproduce the atmospheric muon flux with a good accuracy 
from 1~GeV to 1~TeV, 
the uncertainty of atmospheric neutrino flux due to the
uncertainty of the $\pi$ productions in the interaction model
is estimated from the experimental error and residual of the 
reconstruction of the atmospheric muon flux data.
Summarizing the uncertainties of atmospheric density profile,
interaction cross section and the  $K$-productions in the 
interaction model,
we estimate the uncertainty of the atmospheric neutrino
flux calculated in this paper is 
$\sim$~7\% from 1 to 10~GeV, 
$\sim$~14\% at 100~GeV, and 
$\sim$~25\% at 1~TeV.
Note the statistical error in the Monte Carlo study and uncertainty 
due to the calculation scheme is much smaller than those of 
other sources of the uncertainties.
It is difficult to estimate the error at energies above 1~TeV,
since the uncertainties of the primary flux and the
interaction model would be larger at the corresponding energies 
($\gtrsim$ 10~TeV) of the primary cosmic rays.
Also accurately measured muon flux data are not available
at the corresponding muon momenta ($\gtrsim 3$~TeV)
to calibrate the uncertainties.

The neutrino flux ratios are also compared among the different calculations
to study the variation.
Although the variations for $\nu_\mu/\bar\nu_\mu$ and $\nu_e/\bar\nu_e$
are large, the $(\nu_\mu+\bar\nu_\mu)/(\nu_e+\bar\nu_e)$ ratio is very stable
($\lesssim$ 2\%) over the range 0.1--100~GeV.
The variation of the atmospheric density profile gives the largest
variation in the $(\nu_\mu+\bar\nu_\mu)/(\nu_e+\bar\nu_e)$ ratio.
However, for an one year average, the uncertainty of this ratio will be
sufficiently small.

The stability of the zenith angle dependence is also important
in the study of neutrino oscillations using the atmospheric neutrino flux.
We have studied the ratio [horizontal flux]/[vertical flux] ratio
in the different calculations, then estimated the uncertainty.
The variation or the uncertainty of the ratio is $\sim$~3~\%,
and the variation of $K$-production is considered to be the main
source of this uncertainty.

\begin{acknowledgments}
We greatly appreciate the contributions of J.~Nishimura and A.~Okada
to this paper.
We are grateful to
%K.~Abe, S.~Haino 
%, Y.~Shikaze and S.~Orito
P.G.~Edwards for discussions and comments. 
We also thank the ICRR of the University of Tokyo, 
especially for the use of the computer system.
This study was supported by Grants-in-Aid, KAKENHI(12047206), from
the Ministry of Education, Culture, Sport, Science and Technology 
(MEXT) in Japan.
\end{acknowledgments}

\bibliography{mnflux2}
\clearpage
\appendix
\appendix
\section{Atmospheric neutrino flux below 10 GeV}

Here we tabulate the calculated low energy (0.1--10~GeV) 
atmospheric neutrino flux for 3 locations --- Kamioka, Sudbury (North America), 
and Gran Sasso ---
averaging them in the zenith angle bins with $\Delta \cos\theta_z = 0.1$,
where $\theta_z$ is the zenith angle of the arrival direction
of the neutrino, in Tables II--XX1.

The atmospheric neutrino flux for solar minimum at sea level
could be calculated as
\begin{equation}
\label{eq:flux-value}
\phi_\nu 
= [Value~in~table] \times [Norm] ~~{(\rm m^{-2}sec^{-1}sr^{-1}GeV^{-1})} 
\end{equation}
for each neutrino energy, kind of neutrino, observation site,
and zenith angle bin.

We have also calculated the flux of atmospheric neutrinos for Soudan2
and Frejus sites. The flux at Soudan2 site is almost identical to that of
the SNO site (Sudbury).
The flux at the Frejus site is a little higher 
(3\% at 1~GeV and 10\% at 0.1~GeV)
than that for Gran Sasso. However, the flux for Gran Sasso could be 
used as the approximate flux for the Frejus site.
Therefore, we select the fluxes for Kamioka, Sudbury, and Gran Sasso,
to save space in this paper.

The difference of the fluxes at different sites is due to the
rigidity cutoff of the cosmic rays.
Note, the difference of the terrain above the neutrino detector is not 
considered for each observation site in these tables. It may be
important for the neutrino detectors constructed under high 
mountains.

More detailed flux tables will be available from the web site:  
\url{http://www.icrr.u-tokyo.ac.jp/~mhonda},
from 0.1~GeV to 10~TeV, for solar maximum and solar minimum,
with and without the consideration of the terrain, and
for Kamioka, Sudbury, Soudan2, Gran Sasso, and Frejus sites.
Dividing the azimuthal angels into 12 bins, 
the flux table for each azimuthal bin will also be available
for each calculation conditions stated above.

\begin{table}[tbh]
\caption{Neutrino flux ($\rm m^{-2}sec^{-1}sr^{-1}GeV^{-1}$) for 
$1.0 \ge \cos\theta_z > 0.9$}
\begin{tabular}{c | r r r r| r r r r| r r r r | c}% \hline
\toprule
\multicolumn{1}{c}{}&\multicolumn{4}{|c}{Kamioka}&\multicolumn{4}{|c}{Sudbury}
&\multicolumn{4}{|c}{Gran Sasso}&\multicolumn{1}{|c}{}\\
$E_\nu$ (GeV)&$\nu_\mu$~~&$\bar\nu_\mu$~~&$\nu_e$~~&$\bar\nu_e$~~&$\nu_\mu$~~&$\bar\nu_\mu$~~&$\nu_e$~~&$\bar\nu_e$~~&$\nu_\mu$~~&$\bar\nu_\mu$~~&$\nu_e$~~&$\bar\nu_e$~~&~$Norm$~ ~\\
\toprule
.1000& 35.46& 35.39& 17.88& 17.16& 90.72& 90.33& 49.96& 37.77& 59.58& 59.59& 30.83& 27.54&$10^{2}$\\ 
.1259& 29.84& 29.77& 14.82& 14.08& 74.31& 73.99& 40.25& 30.39& 49.52& 49.54& 25.23& 22.43&$10^{2}$\\ 
.1585& 24.52& 24.39& 11.84& 11.17& 59.41& 58.77& 31.11& 23.53& 40.32& 40.14& 19.95& 17.56&$10^{2}$\\ 
.1995& 19.37& 19.19&  9.13&  8.54& 44.88& 44.37& 22.99& 17.44& 31.40& 31.14& 15.17& 13.24&$10^{2}$\\ 
.2512& 14.49& 14.32&  6.78&  6.25& 31.37& 31.27& 16.14& 12.30& 22.95& 22.77& 11.02&  9.57&$10^{2}$\\ 
.3162& 10.34& 10.16&  4.81&  4.40& 20.54& 20.53& 10.66&  8.20& 15.88& 15.69&  7.62&  6.56&$10^{2}$\\
\botrule
\end{tabular}
\end{table}
 
\begin{table}[tbh]
\caption{Neutrino flux ($\rm m^{-2}sec^{-1}sr^{-1}GeV^{-1}$) for $0.9 \ge \cos\theta_z > 0.8$}
%\caption{Neutrino flux$1.0 \ge \cos\theta_z > 0.9$}
\begin{tabular}{c | r r r r| r r r r| r r r r | c }% \hline
\toprule
\multicolumn{1}{c}{}&\multicolumn{4}{|c}{Kamioka}&\multicolumn{4}{|c}{Sudbury}&\multicolumn{4}{|c}{Gran Sasso}&\multicolumn{1}{|c}{}\\
$E_\nu$ (GeV)&$\nu_\mu$~~&$\bar\nu_\mu$~~&$\nu_e$~~&$\bar\nu_e$~~&$\nu_\mu$~~&$\bar\nu_\mu$~~&$\nu_e$~~&$\bar\nu_e$~~&$\nu_\mu$~~&$\bar\nu_\mu$~~&$\nu_e$~~&$\bar\nu_e$~~&$Norm$~ \\
\toprule
.1000& 36.34& 36.30& 18.31& 17.49& 93.87& 93.35& 51.63& 38.95& 61.25& 61.37& 31.66& 28.28&$10^{2}$\\ 
.1259& 30.34& 30.31& 15.18& 14.31& 76.32& 75.90& 41.46& 31.30& 50.69& 50.69& 25.96& 22.92&$10^{2}$\\ 
.1585& 24.78& 24.68& 12.11& 11.34& 60.55& 60.01& 31.93& 24.15& 41.00& 40.87& 20.44& 17.97&$10^{2}$\\ 
.1995& 19.47& 19.31&  9.32&  8.68& 45.52& 45.07& 23.55& 17.88& 31.77& 31.59& 15.52& 13.58&$10^{2}$\\ 
.2512& 14.52& 14.37&  6.91&  6.39& 31.74& 31.68& 16.56& 12.63& 23.21& 23.04& 11.34&  9.82&$10^{2}$\\ 
.3162& 10.36& 10.21&  4.92&  4.50& 20.80& 20.91& 11.00&  8.45& 16.04& 15.91&  7.88&  6.77&$10^{2}$\\
\botrule
\end{tabular}
\end{table} 
 
\begin{table}[tbh]
\caption{Neutrino flux ($\rm m^{-2}sec^{-1}sr^{-1}GeV^{-1}$) for $0.8 \ge \cos\theta_z> 0.7$}
%\caption{Neutrino flux$1.0 \ge \cos\theta_z > 0.9$}
\begin{tabular}{c | r r r r| r r r r| r r r r | c }% \hline
\toprule
\multicolumn{1}{c}{}&\multicolumn{4}{|c}{Kamioka}&\multicolumn{4}{|c}{Sudbury }&\multicolumn{4}{|c}{Gran Sasso}&\multicolumn{1}{|c}{}\\
$E_\nu$ (GeV)&$\nu_\mu$~~&$\bar\nu_\mu$~~&$\nu_e$~~&$\bar\nu_e$~~&$\nu_\mu$~~&$\bar\nu_\mu$~~&$\nu_e$~~&$\bar\nu_e$~~&$\nu_\mu$~~&$\bar\nu_\mu$~~&$\nu_e$~~&$\bar\nu_e$~~&$Norm$~ \\
\toprule
.1000& 37.43& 37.60& 18.95& 17.91& 97.85& 97.43& 53.65& 40.41& 63.47& 63.64& 32.73& 29.17&$10^{2}$\\ 
.1259& 31.05& 31.09& 15.54& 14.62& 78.82& 78.52& 42.82& 32.39& 52.15& 52.23& 26.65& 23.63&$10^{2}$\\ 
.1585& 25.21& 25.12& 12.37& 11.55& 61.99& 61.52& 32.90& 24.88& 41.78& 41.77& 21.00& 18.48&$10^{2}$\\ 
.1995& 19.67& 19.56&  9.54&  8.83& 46.27& 45.90& 24.23& 18.37& 32.18& 32.05& 15.95& 13.93&$10^{2}$\\ 
.2512& 14.62& 14.51&  7.07&  6.51& 32.14& 32.17& 17.03& 12.99& 23.48& 23.33& 11.61& 10.07&$10^{2}$\\ 
.3162& 10.40& 10.28&  5.04&  4.60& 21.06& 21.24& 11.32&  8.71& 16.19& 16.11&  8.08&  6.95&$10^{2}$\\
\botrule
\end{tabular}
\end{table} 
 
\begin{table}[tbh]
\caption{Neutrino flux ($\rm m^{-2}sec^{-1}sr^{-1}GeV^{-1}$) for $0.7 \ge \cos\theta_z> 0.6$}
%\caption{Neutrino flux$1.0 \ge \cos\theta_z > 0.9$}
\begin{tabular}{c | r r r r| r r r r| r r r r | c }% \hline
\toprule
\multicolumn{1}{c}{}&\multicolumn{4}{|c}{Kamioka}&\multicolumn{4}{|c}{Sudbury }&\multicolumn{4}{|c}{Gran Sasso}&\multicolumn{1}{|c}{}\\
$E_\nu$ (GeV)&$\nu_\mu$~~&$\bar\nu_\mu$~~&$\nu_e$~~&$\bar\nu_e$~~&$\nu_\mu$~~&$\bar\nu_\mu$~~&$\nu_e$~~&$\bar\nu_e$~~&$\nu_\mu$~~&$\bar\nu_\mu$~~&$\nu_e$~~&$\bar\nu_e$~~&$Norm$~ \\
\toprule
.1000& 38.96& 39.24& 19.66& 18.57&103.24&102.73& 56.39& 42.58& 66.52& 66.76& 34.19& 30.51&$10^{2}$\\ 
.1259& 31.96& 32.10& 16.10& 15.08& 82.28& 81.96& 44.72& 33.79& 54.06& 54.26& 27.68& 24.51&$10^{2}$\\ 
.1585& 25.67& 25.72& 12.73& 11.87& 63.95& 63.55& 34.10& 25.82& 42.93& 42.97& 21.66& 18.99&$10^{2}$\\ 
.1995& 19.93& 19.86&  9.76&  9.03& 47.25& 47.00& 24.98& 18.99& 32.79& 32.68& 16.36& 14.26&$10^{2}$\\ 
.2512& 14.78& 14.66&  7.24&  6.62& 32.69& 32.71& 17.50& 13.40& 23.75& 23.63& 11.92& 10.33&$10^{2}$\\ 
.3162& 10.48& 10.37&  5.18&  4.70& 21.38& 21.57& 11.65&  8.98& 16.43& 16.30&  8.31&  7.14&$10^{2}$\\
\botrule
\end{tabular}
\end{table} 
 
\begin{table}[tbh]
\caption{Neutrino flux ($\rm m^{-2}sec^{-1}sr^{-1}GeV^{-1}$) for $0.6 \ge \cos\theta_z> 0.5$}
%\caption{Neutrino flux$1.0 \ge \cos\theta_z > 0.9$}
\begin{tabular}{c | r r r r| r r r r| r r r r | c }% \hline
\toprule
\multicolumn{1}{c}{}&\multicolumn{4}{|c}{Kamioka}&\multicolumn{4}{|c}{Sudbury }&\multicolumn{4}{|c}{Gran Sasso}&\multicolumn{1}{|c}{}\\
$E_\nu$ (GeV)&$\nu_\mu$~~&$\bar\nu_\mu$~~&$\nu_e$~~&$\bar\nu_e$~~&$\nu_\mu$~~&$\bar\nu_\mu$~~&$\nu_e$~~&$\bar\nu_e$~~&$\nu_\mu$~~&$\bar\nu_\mu$~~&$\nu_e$~~&$\bar\nu_e$~~&$Norm$~ \\
\toprule
.1000& 41.34& 41.60& 20.84& 19.58&110.93&110.41& 60.17& 45.48& 71.14& 71.28& 36.16& 32.23&$10^{2}$\\ 
.1259& 33.50& 33.71& 16.84& 15.76& 87.17& 86.97& 47.31& 35.73& 57.05& 57.20& 29.06& 25.74&$10^{2}$\\ 
.1585& 26.58& 26.65& 13.22& 12.24& 66.69& 66.36& 35.75& 27.06& 44.66& 44.62& 22.52& 19.82&$10^{2}$\\ 
.1995& 20.36& 20.36& 10.07&  9.25& 48.62& 48.44& 25.98& 19.76& 33.59& 33.53& 16.91& 14.76&$10^{2}$\\ 
.2512& 14.93& 14.90&  7.45&  6.80& 33.30& 33.48& 18.11& 13.83& 24.10& 24.08& 12.26& 10.60&$10^{2}$\\ 
.3162& 10.58& 10.53&  5.33&  4.80& 21.69& 21.96& 12.02&  9.25& 16.59& 16.54&  8.54&  7.33&$10^{2}$\\
\botrule
\end{tabular}
\end{table} 
 
\begin{table}[tbh]
\caption{Neutrino flux ($\rm m^{-2}sec^{-1}sr^{-1}GeV^{-1}$) for $0.5 \ge \cos\theta_z > 0.4$}
%\caption{Neutrino flux$1.0 \ge \cos\theta_z > 0.9$}
\begin{tabular}{c | r r r r| r r r r| r r r r | c }% \hline
\toprule
\multicolumn{1}{c}{}&\multicolumn{4}{|c}{Kamioka}&\multicolumn{4}{|c}{Sudbury }&\multicolumn{4}{|c}{Gran Sasso}&\multicolumn{1}{|c}{}\\
$E_\nu$ (GeV)&$\nu_\mu$~~&$\bar\nu_\mu$~~&$\nu_e$~~&$\bar\nu_e$~~&$\nu_\mu$~~&$\bar\nu_\mu$~~&$\nu_e$~~&$\bar\nu_e$~~&$\nu_\mu$~~&$\bar\nu_\mu$~~&$\nu_e$~~&$\bar\nu_e$~~&$Norm$~ \\
\toprule
.1000&  4.49&  4.54&  2.25&  2.11& 12.26& 12.16&  6.60&  4.98&  7.78&  7.79&  3.94&  3.51&$10^{3}$\\ 
.1259& 35.90& 36.26& 18.03& 16.69& 94.88& 94.37& 51.25& 38.64& 61.36& 61.44& 31.15& 27.57&$10^{2}$\\ 
.1585& 27.97& 28.19& 13.99& 12.88& 71.11& 70.76& 38.20& 28.89& 47.21& 47.25& 23.91& 20.97&$10^{2}$\\ 
.1995& 21.10& 21.15& 10.56&  9.64& 50.74& 50.71& 27.39& 20.82& 34.96& 34.95& 17.73& 15.46&$10^{2}$\\ 
.2512& 15.30& 15.29&  7.75&  6.98& 34.28& 34.52& 18.88& 14.45& 24.71& 24.70& 12.68& 11.01&$10^{2}$\\ 
.3162& 10.76& 10.71&  5.50&  4.91& 22.20& 22.47& 12.44&  9.63& 16.87& 16.83&  8.78&  7.56&$10^{2}$\\
\botrule
\end{tabular}
\end{table} 
 
\begin{table}[tbh]
\caption{Neutrino flux ($\rm m^{-2}sec^{-1}sr^{-1}GeV^{-1}$) for $0.4 \ge \cos\theta_z > 0.3$}
%\caption{Neutrino flux$1.0 \ge \cos\theta_z > 0.9$}
\begin{tabular}{c | r r r r| r r r r| r r r r | c }% \hline
\toprule
\multicolumn{1}{c}{}&\multicolumn{4}{|c}{Kamioka}&\multicolumn{4}{|c}{Sudbury }&\multicolumn{4}{|c}{Gran Sasso}&\multicolumn{1}{|c}{}\\
$E_\nu$ (GeV)&$\nu_\mu$~~&$\bar\nu_\mu$~~&$\nu_e$~~&$\bar\nu_e$~~&$\nu_\mu$~~&$\bar\nu_\mu$~~&$\nu_e$~~&$\bar\nu_e$~~&$\nu_\mu$~~&$\bar\nu_\mu$~~&$\nu_e$~~&$\bar\nu_e$~~&$Norm$~ \\
\toprule
.1000&  5.10&  5.16&  2.54&  2.37& 14.14& 14.03&  7.56&  5.68&  8.88&  8.91&  4.46&  3.97&$10^{3}$\\ 
.1259& 39.93& 40.41& 20.02& 18.52&107.11&106.64& 57.68& 43.54& 68.83& 68.89& 34.82& 30.77&$10^{2}$\\ 
.1585& 30.50& 30.69& 15.28& 13.97& 78.28& 78.02& 42.24& 31.96& 51.63& 51.71& 26.12& 22.99&$10^{2}$\\ 
.1995& 22.50& 22.61& 11.37& 10.27& 54.48& 54.68& 29.70& 22.65& 37.39& 37.35& 19.00& 16.63&$10^{2}$\\ 
.2512& 15.98& 16.11&  8.22&  7.37& 36.00& 36.47& 20.05& 15.43& 25.98& 25.91& 13.45& 11.64&$10^{2}$\\ 
.3162& 11.05& 11.06&  5.74&  5.10& 22.92& 23.33& 13.04& 10.12& 17.35& 17.34&  9.19&  7.90&$10^{2}$\\
\botrule
\end{tabular}
\end{table} 
 
\begin{table}[tbh]
\caption{Neutrino flux ($\rm m^{-2}sec^{-1}sr^{-1}GeV^{-1}$) for $0.3 \ge \cos\theta_z > 0.2$}
%\caption{Neutrino flux$1.0 \ge \cos\theta_z > 0.9$}
\begin{tabular}{c | r r r r| r r r r| r r r r | c }% \hline
\toprule
\multicolumn{1}{c}{}&\multicolumn{4}{|c}{Kamioka}&\multicolumn{4}{|c}{Sudbury }&\multicolumn{4}{|c}{Gran Sasso}&\multicolumn{1}{|c}{}\\
$E_\nu$ (GeV)&$\nu_\mu$~~&$\bar\nu_\mu$~~&$\nu_e$~~&$\bar\nu_e$~~&$\nu_\mu$~~&$\bar\nu_\mu$~~&$\nu_e$~~&$\bar\nu_e$~~&$\nu_\mu$~~&$\bar\nu_\mu$~~&$\nu_e$~~&$\bar\nu_e$~~&$Norm$~ \\
\toprule
.1000&  6.23&  6.30&  3.08&  2.86& 17.48& 17.38&  9.28&  6.98& 10.91& 10.93&  5.41&  4.85&$10^{3}$\\ 
.1259&  4.76&  4.82&  2.38&  2.19& 12.95& 12.93&  6.96&  5.23&  8.25&  8.25&  4.13&  3.66&$10^{3}$\\ 
.1585& 35.35& 35.73& 17.81& 16.13& 91.76& 92.00& 49.75& 37.59& 60.22& 60.32& 30.42& 26.74&$10^{2}$\\ 
.1995& 25.28& 25.54& 12.92& 11.56& 61.70& 62.21& 34.07& 25.97& 42.25& 42.33& 21.64& 18.87&$10^{2}$\\ 
.2512& 17.43& 17.63&  9.11&  8.07& 39.57& 40.15& 22.44& 17.30& 28.40& 28.48& 14.90& 12.88&$10^{2}$\\ 
.3162& 11.75& 11.85&  6.28&  5.48& 24.59& 25.07& 14.19& 11.10& 18.51& 18.56&  9.97&  8.57&$10^{2}$\\
\botrule
\end{tabular}
\end{table} 
 
\begin{table}[tbh]
\caption{Neutrino flux ($\rm m^{-2}sec^{-1}sr^{-1}GeV^{-1}$) for $0.2 \ge \cos\theta_z > 0.1$}
%\caption{Neutrino flux$1.0 \ge \cos\theta_z > 0.9$}
\begin{tabular}{c | r r r r| r r r r| r r r r | c }% \hline
\toprule
\multicolumn{1}{c}{}&\multicolumn{4}{|c}{Kamioka}&\multicolumn{4}{|c}{Sudbury }&\multicolumn{4}{|c}{Gran Sasso}&\multicolumn{1}{|c}{}\\
$E_\nu$ (GeV)&$\nu_\mu$~~&$\bar\nu_\mu$~~&$\nu_e$~~&$\bar\nu_e$~~&$\nu_\mu$~~&$\bar\nu_\mu$~~&$\nu_e$~~&$\bar\nu_e$~~&$\nu_\mu$~~&$\bar\nu_\mu$~~&$\nu_e$~~&$\bar\nu_e$~~&$Norm$~ \\
\toprule
.1000&  8.66&  8.76&  4.23&  3.92& 24.56& 24.41& 12.94&  9.70& 15.19& 15.25&  7.46&  6.67&$10^{3}$\\ 
.1259&  6.46&  6.52&  3.19&  2.93& 17.73& 17.69&  9.47&  7.12& 11.24& 11.23&  5.58&  4.93&$10^{3}$\\ 
.1585&  4.62&  4.68&  2.33&  2.09& 12.11& 12.17&  6.59&  4.97&  7.92&  7.93&  3.99&  3.51&$10^{3}$\\ 
.1995& 31.75& 32.26& 16.38& 14.56& 78.09& 79.26& 43.65& 33.38& 53.16& 53.57& 27.59& 24.01&$10^{2}$\\ 
.2512& 21.09& 21.42& 11.19&  9.86& 47.95& 49.10& 27.65& 21.50& 34.40& 34.61& 18.43& 15.82&$10^{2}$\\ 
.3162& 13.69& 13.92&  7.43&  6.44& 28.67& 29.46& 16.84& 13.30& 21.58& 21.70& 11.80& 10.11&$10^{2}$\\
\botrule
\end{tabular}
\end{table} 
\clearpage
 
\begin{table}[tbh]
\caption{Neutrino flux ($\rm m^{-2}sec^{-1}sr^{-1}GeV^{-1}$) for $0.1 \ge \cos\theta_z> 0.0$}
%\caption{Neutrino flux$1.0 \ge \cos\theta_z > 0.9$}
\begin{tabular}{c | r r r r| r r r r| r r r r | c }% \hline
\toprule
\multicolumn{1}{c}{}&\multicolumn{4}{|c}{Kamioka}&\multicolumn{4}{|c}{Sudbury }&\multicolumn{4}{|c}{Gran Sasso}&\multicolumn{1}{|c}{}\\
$E_\nu$ (GeV)&$\nu_\mu$~~&$\bar\nu_\mu$~~&$\nu_e$~~&$\bar\nu_e$~~&$\nu_\mu$~~&$\bar\nu_\mu$~~&$\nu_e$~~&$\bar\nu_e$~~&$\nu_\mu$~~&$\bar\nu_\mu$~~&$\nu_e$~~&$\bar\nu_e$~~&$Norm$~ \\
\toprule
.1000& 14.19& 14.40&  6.85&  6.38& 39.91& 39.56& 20.65& 15.71& 24.89& 24.87& 12.08& 10.70&$10^{3}$\\ 
.1259& 10.27& 10.43&  5.01&  4.62& 28.11& 27.96& 14.85& 11.23& 17.85& 17.85&  8.78&  7.77&$10^{3}$\\ 
.1585&  7.10&  7.19&  3.54&  3.21& 18.41& 18.60& 10.01&  7.59& 12.09& 12.13&  6.12&  5.35&$10^{3}$\\ 
.1995&  4.68&  4.73&  2.40&  2.14& 11.30& 11.60&  6.41&  4.90&  7.75&  7.80&  4.07&  3.52&$10^{3}$\\ 
.2512& 29.50& 30.04& 15.88& 13.70& 66.33& 68.44& 39.32& 30.33& 47.63& 47.94& 25.83& 22.23&$10^{2}$\\ 
.3162& 18.20& 18.53& 10.09&  8.67& 37.91& 39.19& 22.90& 18.05& 28.51& 28.77& 16.00& 13.60&$10^{2}$\\
\botrule
\end{tabular}
\end{table} 
 
\begin{table}[tbh]
\caption{Neutrino flux ($\rm m^{-2}sec^{-1}sr^{-1}GeV^{-1}$) for $0.0 \ge \cos\theta_z> -0.1$}
%\caption{Neutrino flux$1.0 \ge \cos\theta_z > 0.9$}
\begin{tabular}{c | r r r r| r r r r| r r r r | c }% \hline
\toprule
\multicolumn{1}{c}{}&\multicolumn{4}{|c}{Kamioka}&\multicolumn{4}{|c}{Sudbury }&\multicolumn{4}{|c}{Gran Sasso}&\multicolumn{1}{|c}{}\\
$E_\nu$ (GeV)&$\nu_\mu$~~&$\bar\nu_\mu$~~&$\nu_e$~~&$\bar\nu_e$~~&$\nu_\mu$~~&$\bar\nu_\mu$~~&$\nu_e$~~&$\bar\nu_e$~~&$\nu_\mu$~~&$\bar\nu_\mu$~~&$\nu_e$~~&$\bar\nu_e$~~&$Norm$~ \\
\toprule
.1000& 15.16& 15.36&  7.33&  6.74& 38.57& 38.36& 19.77& 15.33& 24.81& 24.96& 12.23& 10.64&$10^{3}$\\ 
.1259& 11.02& 11.06&  5.33&  4.89& 27.25& 27.11& 14.22& 11.01& 17.84& 17.89&  8.92&  7.72&$10^{3}$\\ 
.1585&  7.55&  7.59&  3.76&  3.39& 17.94& 18.07&  9.68&  7.50& 12.09& 12.16&  6.16&  5.28&$10^{3}$\\ 
.1995&  4.91&  4.96&  2.54&  2.26& 11.12& 11.34&  6.24&  4.85&  7.70&  7.78&  4.07&  3.45&$10^{3}$\\ 
.2512& 30.78& 31.12& 16.47& 14.43& 65.79& 67.71& 38.40& 30.08& 46.67& 47.52& 25.89& 21.83&$10^{2}$\\ 
.3162& 18.71& 19.04& 10.41&  8.96& 37.76& 39.04& 22.57& 18.03& 27.81& 28.35& 15.88& 13.31&$10^{2}$\\
\botrule
\end{tabular}
\end{table} 

\begin{table}[tbh]
\caption{Neutrino flux ($\rm m^{-2}sec^{-1}sr^{-1}GeV^{-1}$) for $-0.1 \ge \cos\theta_z > -0.2$}
%\caption{Neutrino flux$1.0 \ge \cos\theta_z > 0.9$}
\begin{tabular}{c | r r r r| r r r r| r r r r | c }% \hline
\toprule
\multicolumn{1}{c}{}&\multicolumn{4}{|c}{Kamioka}&\multicolumn{4}{|c}{Sudbury }&\multicolumn{4}{|c}{Gran Sasso}&\multicolumn{1}{|c}{}\\
$E_\nu$ (GeV)&$\nu_\mu$~~&$\bar\nu_\mu$~~&$\nu_e$~~&$\bar\nu_e$~~&$\nu_\mu$~~&$\bar\nu_\mu$~~&$\nu_e$~~&$\bar\nu_e$~~&$\nu_\mu$~~&$\bar\nu_\mu$~~&$\nu_e$~~&$\bar\nu_e$~~&$Norm$~ \\
\toprule
.1000& 10.87& 10.94&  5.36&  4.76& 21.25& 21.17& 10.95&  8.70& 15.11& 15.14&  7.68&  6.37&$10^{3}$\\ 
.1259&  8.00&  8.03&  3.99&  3.51& 15.49& 15.48&  8.09&  6.42& 11.05& 11.11&  5.70&  4.69&$10^{3}$\\ 
.1585&  5.61&  5.66&  2.85&  2.50& 10.71& 10.75&  5.70&  4.53&  7.70&  7.76&  4.04&  3.31&$10^{3}$\\ 
.1995& 37.64& 38.05& 19.62& 17.06& 70.25& 71.21& 38.50& 30.55& 51.06& 51.75& 27.49& 22.55&$10^{2}$\\ 
.2512& 24.34& 24.56& 13.07& 11.24& 44.20& 45.09& 24.94& 19.91& 32.48& 33.15& 18.01& 14.83&$10^{2}$\\ 
.3162& 15.28& 15.46&  8.41&  7.17& 27.06& 27.62& 15.58& 12.54& 20.15& 20.61& 11.43&  9.43&$10^{2}$\\
\botrule
\end{tabular}
\end{table} 

\begin{table}[tbh]
\caption{Neutrino flux ($\rm m^{-2}sec^{-1}sr^{-1}GeV^{-1}$) for $-0.2 \ge \cos\theta_z > -0.3$}
%\caption{Neutrino flux$1.0 \ge \cos\theta_z > 0.9$}
\begin{tabular}{c | r r r r| r r r r| r r r r | c }% \hline
\toprule
\multicolumn{1}{c}{}&\multicolumn{4}{|c}{Kamioka}&\multicolumn{4}{|c}{Sudbury }&\multicolumn{4}{|c}{Gran Sasso}&\multicolumn{1}{|c}{}\\
$E_\nu$ (GeV)&$\nu_\mu$~~&$\bar\nu_\mu$~~&$\nu_e$~~&$\bar\nu_e$~~&$\nu_\mu$~~&$\bar\nu_\mu$~~&$\nu_e$~~&$\bar\nu_e$~~&$\nu_\mu$~~&$\bar\nu_\mu$~~&$\nu_e$~~&$\bar\nu_e$~~&$Norm$~ \\
\toprule
.1000&  8.65&  8.65&  4.41&  3.71& 12.70& 12.71&  6.59&  5.30& 10.35& 10.37&  5.34&  4.36&$10^{3}$\\ 
.1259&  6.49&  6.50&  3.35&  2.80&  9.52&  9.54&  4.99&  4.01&  7.77&  7.80&  4.06&  3.29&$10^{3}$\\ 
.1585&  4.68&  4.70&  2.44&  2.04&  6.87&  6.91&  3.64&  2.91&  5.60&  5.64&  2.96&  2.39&$10^{3}$\\ 
.1995& 32.35& 32.56& 17.09& 14.30& 47.44& 47.87& 25.52& 20.46& 38.70& 39.10& 20.76& 16.82&$10^{2}$\\ 
.2512& 21.46& 21.64& 11.59&  9.71& 31.32& 31.76& 17.19& 13.94& 25.67& 26.09& 14.07& 11.44&$10^{2}$\\ 
.3162& 13.89& 14.02&  7.63&  6.41& 20.09& 20.38& 11.26&  9.16& 16.53& 16.84&  9.20&  7.56&$10^{2}$\\
\botrule
\end{tabular}
\end{table} 
 
\begin{table}[tbh]
\caption{Neutrino flux ($\rm m^{-2}sec^{-1}sr^{-1}GeV^{-1}$) for $-0.3 \ge \cos\theta_z > -0.4$}
%\caption{Neutrino flux$1.0 \ge \cos\theta_z > 0.9$}
\begin{tabular}{c | r r r r| r r r r| r r r r | c }% \hline
\toprule
\multicolumn{1}{c}{}&\multicolumn{4}{|c}{Kamioka}&\multicolumn{4}{|c}{Sudbury }&\multicolumn{4}{|c}{Gran Sasso}&\multicolumn{1}{|c}{}\\
$E_\nu$ (GeV)&$\nu_\mu$~~&$\bar\nu_\mu$~~&$\nu_e$~~&$\bar\nu_e$~~&$\nu_\mu$~~&$\bar\nu_\mu$~~&$\nu_e$~~&$\bar\nu_e$~~&$\nu_\mu$~~&$\bar\nu_\mu$~~&$\nu_e$~~&$\bar\nu_e$~~&$Norm$~ \\ 
\toprule
.1000&  7.18&  7.17&  3.71&  3.07&  8.78&  8.78&  4.58&  3.73&  8.11&  8.10&  4.24&  3.44&$10^{3}$\\ 
.1259&  5.50&  5.50&  2.86&  2.37&  6.74&  6.76&  3.55&  2.88&  6.22&  6.23&  3.27&  2.66&$10^{3}$\\ 
.1585& 40.89& 40.91& 21.30& 17.66& 50.16& 50.40& 26.41& 21.47& 46.30& 46.41& 24.36& 19.70&$10^{2}$\\ 
.1995& 29.09& 29.20& 15.31& 12.68& 35.81& 36.09& 18.98& 15.46& 32.97& 33.19& 17.52& 14.22&$10^{2}$\\ 
.2512& 19.83& 20.01& 10.61&  8.83& 24.43& 24.74& 13.20& 10.78& 22.41& 22.73& 12.15&  9.95&$10^{2}$\\ 
.3162& 13.17& 13.26&  7.13&  5.95& 16.12& 16.36&  8.88&  7.27& 14.82& 15.04&  8.16&  6.67&$10^{2}$\\
\botrule
\end{tabular}
\end{table} 
 
\begin{table}[tbh]
\caption{Neutrino flux ($\rm m^{-2}sec^{-1}sr^{-1}GeV^{-1}$) for $-0.4 \ge \cos\theta_z > -0.5$}
%\caption{Neutrino flux$1.0 \ge \cos\theta_z > 0.9$}
\begin{tabular}{c | r r r r| r r r r| r r r r | c }% \hline
\toprule
\multicolumn{1}{c}{}&\multicolumn{4}{|c}{Kamioka}&\multicolumn{4}{|c}{Sudbury }&\multicolumn{4}{|c}{Gran Sasso}&\multicolumn{1}{|c}{}\\
$E_\nu$ (GeV)&$\nu_\mu$~~&$\bar\nu_\mu$~~&$\nu_e$~~&$\bar\nu_e$~~&$\nu_\mu$~~&$\bar\nu_\mu$~~&$\nu_e$~~&$\bar\nu_e$~~&$\nu_\mu$~~&$\bar\nu_\mu$~~&$\nu_e$~~&$\bar\nu_e$~~&$Norm$~ \\
\toprule
.1000&  6.38&  6.38&  3.32&  2.77&  6.82&  6.81&  3.57&  2.96&  6.92&  6.92&  3.64&  2.98&$10^{3}$\\ 
.1259&  5.00&  5.00&  2.61&  2.18&  5.35&  5.36&  2.80&  2.32&  5.42&  5.44&  2.86&  2.34&$10^{3}$\\ 
.1585& 38.12& 38.05& 19.76& 16.47& 40.84& 40.86& 21.25& 17.61& 41.36& 41.44& 21.68& 17.72&$10^{2}$\\ 
.1995& 27.88& 27.83& 14.45& 12.08& 29.94& 29.95& 15.61& 12.94& 30.21& 30.29& 15.87& 13.00&$10^{2}$\\ 
.2512& 19.43& 19.48& 10.21&  8.58& 20.90& 21.01& 11.08&  9.18& 20.99& 21.18& 11.19&  9.21&$10^{2}$\\ 
.3162& 13.07& 13.11&  6.99&  5.87& 14.07& 14.16&  7.56&  6.27& 14.11& 14.26&  7.64&  6.28&$10^{2}$\\
\botrule
\end{tabular}
\end{table} 
 
\begin{table}[tbh]
\caption{Neutrino flux ($\rm m^{-2}sec^{-1}sr^{-1}GeV^{-1}$) for $-0.5 \ge \cos\theta_z > -0.6$}
%\caption{Neutrino flux$1.0 \ge \cos\theta_z > 0.9$}
\begin{tabular}{c | r r r r| r r r r| r r r r | c }% \hline
\toprule
\multicolumn{1}{c}{}&\multicolumn{4}{|c}{Kamioka}&\multicolumn{4}{|c}{Sudbury }&\multicolumn{4}{|c}{Gran Sasso}&\multicolumn{1}{|c}{}\\
$E_\nu$ (GeV)&$\nu_\mu$~~&$\bar\nu_\mu$~~&$\nu_e$~~&$\bar\nu_e$~~&$\nu_\mu$~~&$\bar\nu_\mu$~~&$\nu_e$~~&$\bar\nu_e$~~&$\nu_\mu$~~&$\bar\nu_\mu$~~&$\nu_e$~~&$\bar\nu_e$~~&$Norm$~ \\
\toprule
.1000&  6.34&  6.33&  3.30&  2.78&  5.61&  5.62&  2.93&  2.49&  6.38&  6.41&  3.36&  2.81&$10^{3}$\\ 
.1259&  5.04&  5.04&  2.63&  2.20&  4.48&  4.49&  2.33&  1.99&  5.09&  5.11&  2.67&  2.23&$10^{3}$\\ 
.1585& 39.19& 39.05& 20.17& 16.92& 34.95& 34.89& 17.96& 15.29& 39.67& 39.66& 20.62& 17.07&$10^{2}$\\ 
.1995& 29.21& 29.07& 14.97& 12.57& 26.20& 26.15& 13.39& 11.38& 29.67& 29.62& 15.35& 12.70&$10^{2}$\\ 
.2512& 20.67& 20.62& 10.70&  8.99& 18.66& 18.70&  9.63&  8.17& 21.02& 21.05& 10.98&  9.12&$10^{2}$\\ 
.3162& 14.06& 14.02&  7.36&  6.18& 12.79& 12.78&  6.68&  5.66& 14.26& 14.32&  7.55&  6.27&$10^{2}$\\
\botrule
\end{tabular}
\end{table} 
 
\begin{table}[tbh]
\caption{Neutrino flux ($\rm m^{-2}sec^{-1}sr^{-1}GeV^{-1}$) for $-0.6 \ge \cos\theta_z > -0.7$}
%\caption{Neutrino flux$1.0 \ge \cos\theta_z > 0.9$}
\begin{tabular}{c | r r r r| r r r r| r r r r | c }% \hline
\toprule
\multicolumn{1}{c}{}&\multicolumn{4}{|c}{Kamioka}&\multicolumn{4}{|c}{Sudbury }&\multicolumn{4}{|c}{Gran Sasso}&\multicolumn{1}{|c}{}\\
$E_\nu$ (GeV)&$\nu_\mu$~~&$\bar\nu_\mu$~~&$\nu_e$~~&$\bar\nu_e$~~&$\nu_\mu$~~&$\bar\nu_\mu$~~&$\nu_e$~~&$\bar\nu_e$~~&$\nu_\mu$~~&$\bar\nu_\mu$~~&$\nu_e$~~&$\bar\nu_e$~~&$Norm$~ \\
\toprule
.1000&  6.59&  6.58&  3.49&  2.87&  4.70&  4.71&  2.41&  2.17&  6.04&  6.05&  3.16&  2.69&$10^{3}$\\ 
.1259&  5.31&  5.30&  2.79&  2.30&  3.83&  3.84&  1.96&  1.76&  4.89&  4.90&  2.55&  2.18&$10^{3}$\\ 
.1585& 41.77& 41.56& 21.59& 17.76& 30.47& 30.44& 15.32& 13.64& 38.71& 38.75& 19.86& 16.85&$10^{2}$\\ 
.1995& 31.43& 31.23& 16.11& 13.25& 23.33& 23.23& 11.61& 10.28& 29.43& 29.41& 14.95& 12.61&$10^{2}$\\ 
.2512& 22.33& 22.30& 11.54&  9.51& 17.01& 16.93&  8.50&  7.51& 21.20& 21.17& 10.82&  9.09&$10^{2}$\\ 
.3162& 15.10& 15.14&  7.89&  6.51& 11.85& 11.81&  5.98&  5.25& 14.52& 14.50&  7.49&  6.27&$10^{2}$\\
\botrule
\end{tabular}
\end{table} 
 
\begin{table}[tbh]
\caption{Neutrino flux ($\rm m^{-2}sec^{-1}sr^{-1}GeV^{-1}$) for $-0.7 \ge \cos\theta_z > -0.8$}
%\caption{Neutrino flux$1.0 \ge \cos\theta_z > 0.9$}
\begin{tabular}{c | r r r r| r r r r| r r r r | c }% \hline
\toprule
\multicolumn{1}{c}{}&\multicolumn{4}{|c}{Kamioka}&\multicolumn{4}{|c}{Sudbury }&\multicolumn{4}{|c}{Gran Sasso}&\multicolumn{1}{|c}{}\\
$E_\nu$ (GeV)&$\nu_\mu$~~&$\bar\nu_\mu$~~&$\nu_e$~~&$\bar\nu_e$~~&$\nu_\mu$~~&$\bar\nu_\mu$~~&$\nu_e$~~&$\bar\nu_e$~~&$\nu_\mu$~~&$\bar\nu_\mu$~~&$\nu_e$~~&$\bar\nu_e$~~&$Norm$~ \\
\toprule
.1000&  6.27&  6.26&  3.32&  2.76&  4.44&  4.43&  2.27&  2.08&  5.35&  5.37&  2.80&  2.43&$10^{3}$\\ 
.1259&  5.10&  5.10&  2.68&  2.22&  3.66&  3.66&  1.86&  1.69&  4.40&  4.41&  2.29&  1.97&$10^{3}$\\ 
.1585& 40.67& 40.48& 20.90& 17.28& 29.57& 29.45& 14.71& 13.29& 35.25& 35.23& 17.97& 15.36&$10^{2}$\\ 
.1995& 30.96& 30.77& 15.69& 12.96& 22.95& 22.77& 11.24& 10.09& 27.09& 27.00& 13.60& 11.60&$10^{2}$\\ 
.2512& 22.18& 22.11& 11.28&  9.32& 16.90& 16.74&  8.27&  7.37& 19.67& 19.60&  9.88&  8.41&$10^{2}$\\ 
.3162& 15.07& 15.03&  7.72&  6.40& 11.88& 11.75&  5.84&  5.16& 13.55& 13.53&  6.85&  5.79&$10^{2}$\\
\botrule
\end{tabular}
\end{table}
 
\begin{table}[tbh]
\caption{Neutrino flux ($\rm m^{-2}sec^{-1}sr^{-1}GeV^{-1}$) for $-0.8 \ge \cos\theta_z >-0.9$}
%\caption{Neutrino flux$1.0 \ge \cos\theta_z > 0.9$}
\begin{tabular}{c | r r r r| r r r r| r r r r | c }% \hline
\toprule
\multicolumn{1}{c}{}&\multicolumn{4}{|c}{Kamioka}&\multicolumn{4}{|c}{Sudbury }&\multicolumn{4}{|c}{Gran Sasso}&\multicolumn{1}{|c}{}\\
$E_\nu$ (GeV)&$\nu_\mu$~~&$\bar\nu_\mu$~~&$\nu_e$~~&$\bar\nu_e$~~&$\nu_\mu$~~&$\bar\nu_\mu$~~&$\nu_e$~~&$\bar\nu_e$~~&$\nu_\mu$~~&$\bar\nu_\mu$~~&$\nu_e$~~&$\bar\nu_e$~~&$Norm$~ \\
\toprule
.1000&  5.54&  5.54&  2.94&  2.47&  5.14&  5.12&  2.70&  2.30&  5.09&  5.08&  2.69&  2.27&$10^{3}$\\ 
.1259&  4.57&  4.56&  2.39&  2.01&  4.23&  4.22&  2.20&  1.86&  4.19&  4.19&  2.19&  1.85&$10^{3}$\\ 
.1585& 36.78& 36.55& 18.76& 15.71& 34.07& 33.78& 17.26& 14.57& 33.82& 33.61& 17.20& 14.46&$10^{2}$\\ 
.1995& 28.27& 28.07& 14.15& 11.82& 26.20& 25.90& 13.02& 10.97& 26.07& 25.83& 12.99& 10.91&$10^{2}$\\ 
.2512& 20.41& 20.32& 10.19&  8.52& 18.96& 18.76&  9.40&  7.92& 18.89& 18.77&  9.38&  7.90&$10^{2}$\\ 
.3162& 13.97& 13.90&  7.00&  5.86& 12.97& 12.87&  6.46&  5.45& 12.97& 12.87&  6.46&  5.43&$10^{2}$\\
\botrule
\end{tabular}
\end{table} 
 
\begin{table}[tbh]
\caption{Neutrino flux ($\rm m^{-2}sec^{-1}sr^{-1}GeV^{-1}$) for $-0.9 \ge \cos\theta_z \ge -1.0$}
%\caption{Neutrino flux$1.0 \ge \cos\theta_z > 0.9$}
\begin{tabular}{c | r r r r| r r r r| r r r r | c }% \hline
\toprule
\multicolumn{1}{c}{}&\multicolumn{4}{|c}{Kamioka}&\multicolumn{4}{|c}{Sudbury }&\multicolumn{4}{|c}{Gran Sasso}&\multicolumn{1}{|c}{}\\
$E_\nu$ (GeV)&$\nu_\mu$~~&$\bar\nu_\mu$~~&$\nu_e$~~&$\bar\nu_e$~~&$\nu_\mu$~~&$\bar\nu_\mu$~~&$\nu_e$~~&$\bar\nu_e$~~&$\nu_\mu$~~&$\bar\nu_\mu$~~&$\nu_e$~~&$\bar\nu_e$~~&$Norm$~ \\
\toprule
.1000&  5.36&  5.36&  2.82&  2.41&  6.51&  6.49&  3.49&  2.83&  5.82&  5.81&  3.11&  2.57&$10^{3}$\\ 
.1259& 44.49& 44.41& 23.09& 19.67& 53.59& 53.54& 28.38& 22.91& 48.10& 47.97& 25.22& 20.90&$10^{2}$\\ 
.1585& 36.08& 35.90& 18.15& 15.43& 43.23& 42.80& 22.09& 17.86& 38.90& 38.63& 19.75& 16.29&$10^{2}$\\ 
.1995& 27.96& 27.73& 13.72& 11.65& 33.09& 32.63& 16.50& 13.36& 29.95& 29.65& 14.84& 12.22&$10^{2}$\\ 
.2512& 20.35& 20.18&  9.94&  8.41& 23.57& 23.38& 11.77&  9.53& 21.52& 21.34& 10.63&  8.78&$10^{2}$\\ 
.3162& 13.98& 13.83&  6.83&  5.75& 15.83& 15.75&  7.94&  6.44& 14.60& 14.48&  7.23&  5.97&$10^{2}$\\
\botrule
\end{tabular}
\end{table} 

\clearpage
\section{Atmospheric neutrino flux above 10~GeV}

Here we tabulate the atmospheric neutrino flux calculated in this work
for neutrino energies above 10~GeV in Tables XXII--XXV.
The atmospheric neutrino flux could also be calculated 
using Eq. \ref{eq:flux-value}.

Note, we tabulate one kind of neutrino flux in one table independently 
of the observation site, for the down going directions.
The neutrino flux for upward going direction is obtained using the 
mirror symmetry of the atmospheric neutrino flux:
\begin{equation}
\phi_\nu (-\cos\theta) = \phi_\nu (\cos\theta) 
\end{equation}
valid in the energy region where 
the rigidity cutoff does not affect the neutrino flux.

\begin{table}[tbh]
\caption{$\nu_\mu$ flux ($\rm m^{-2}sec^{-1}sr^{-1}GeV^{-1}$) above 10~GeV}
\end{table} 
\begin{supertabular}{c | r r r r r r r r r r | c }% \hline
\toprule
\multicolumn{1}{c|}{}&\multicolumn{10}{c}{$\cos\theta_z$}&\multicolumn{1}{|c}{}\\
$E_\nu$~(GeV)~&~1.--.9~&~.9--.8~&~.8--.7~&~.7--.6~&~.6--.5~&~.5--.4~&~.4--.3~&~.3--.2~&~.2--.1~&~.1--.0~&$Norm$\\
\toprule
 1.000$\times 10^{1}$~&~2.557~&~2.625~&~2.703~&~2.799~&~2.911~&~3.052~&~3.232~&~3.482~&~3.824~&~4.172~&~$10^{-1}$\\
 1.259$\times 10^{1}$~&~1.295~&~1.331~&~1.370~&~1.419~&~1.479~&~1.554~&~1.646~&~1.778~&~1.964~&~2.166~&~$10^{-1}$\\
 1.585$\times 10^{1}$~&~0.654~&~0.673~&~0.694~&~0.720~&~0.751~&~0.789~&~0.840~&~0.906~&~1.009~&~1.121~&~$10^{-1}$\\
 1.995$\times 10^{1}$~&~3.297~&~3.397~&~3.505~&~3.653~&~3.811~&~4.001~&~4.269~&~4.612~&~5.154~&~5.807~&~$10^{-2}$\\
 2.512$\times 10^{1}$~&~1.659~&~1.710~&~1.770~&~1.848~&~1.930~&~2.033~&~2.167~&~2.349~&~2.627~&~2.997~&~$10^{-2}$\\
 3.162$\times 10^{1}$~&~0.831~&~0.858~&~0.891~&~0.931~&~0.974~&~1.033~&~1.100~&~1.197~&~1.340~&~1.542~&~$10^{-2}$\\
 3.981$\times 10^{1}$~&~4.144~&~4.291~&~4.463~&~4.663~&~4.898~&~5.205~&~5.572~&~6.091~&~6.852~&~7.935~&~$10^{-3}$\\
 5.012$\times 10^{1}$~&~2.055~&~2.136~&~2.225~&~2.329~&~2.457~&~2.612~&~2.819~&~3.085~&~3.482~&~4.051~&~$10^{-3}$\\
 6.310$\times 10^{1}$~&~1.014~&~1.056~&~1.104~&~1.161~&~1.228~&~1.308~&~1.420~&~1.556~&~1.762~&~2.059~&~$10^{-3}$\\
 7.943$\times 10^{1}$~&~0.499~&~0.519~&~0.545~&~0.576~&~0.609~&~0.653~&~0.710~&~0.783~&~0.897~&~1.054~&~$10^{-3}$\\
 1.000$\times 10^{2}$~&~2.443~&~2.551~&~2.679~&~2.838~&~3.012~&~3.248~&~3.541~&~3.930~&~4.524~&~5.345~&~$10^{-4}$\\
 1.259$\times 10^{2}$~&~1.194~&~1.253~&~1.315~&~1.394~&~1.487~&~1.606~&~1.761~&~1.967~&~2.259~&~2.676~&~$10^{-4}$\\
 1.585$\times 10^{2}$~&~0.583~&~0.611~&~0.643~&~0.684~&~0.732~&~0.790~&~0.869~&~0.979~&~1.129~&~1.338~&~$10^{-4}$\\
 1.995$\times 10^{2}$~&~2.837~&~2.969~&~3.134~&~3.340~&~3.568~&~3.876~&~4.270~&~4.843~&~5.619~&~6.676~&~$10^{-5}$\\
 2.512$\times 10^{2}$~&~1.371~&~1.439~&~1.521~&~1.621~&~1.732~&~1.897~&~2.092~&~2.384~&~2.785~&~3.322~&~$10^{-5}$\\
 3.162$\times 10^{2}$~&~0.658~&~0.695~&~0.737~&~0.786~&~0.844~&~0.923~&~1.022~&~1.168~&~1.378~&~1.646~&~$10^{-5}$\\
 3.981$\times 10^{2}$~&~3.146~&~3.328~&~3.547~&~3.792~&~4.096~&~4.482~&~4.988~&~5.700~&~6.771~&~8.124~&~$10^{-6}$\\
 5.012$\times 10^{2}$~&~1.496~&~1.585~&~1.696~&~1.819~&~1.975~&~2.171~&~2.425~&~2.776~&~3.308~&~3.990~&~$10^{-6}$\\
 6.310$\times 10^{2}$~&~0.706~&~0.753~&~0.806~&~0.869~&~0.949~&~1.045~&~1.172~&~1.353~&~1.617~&~1.950~&~$10^{-6}$\\
 7.943$\times 10^{2}$~&~3.307~&~3.537~&~3.807~&~4.123~&~4.521~&~5.008~&~5.643~&~6.568~&~7.855~&~9.512~&~$10^{-7}$\\
 1.000$\times 10^{3}$~&~1.535~&~1.643~&~1.781~&~1.940~&~2.133~&~2.386~&~2.708~&~3.167~&~3.796~&~4.634~&~$10^{-7}$\\
 1.259$\times 10^{3}$~&~0.705~&~0.759~&~0.825~&~0.905~&~1.001~&~1.125~&~1.288~&~1.515~&~1.840~&~2.250~&~$10^{-7}$\\
 1.585$\times 10^{3}$~&~0.320~&~0.347~&~0.378~&~0.418~&~0.465~&~0.526~&~0.608~&~0.722~&~0.886~&~1.088~&~$10^{-7}$\\
 1.995$\times 10^{3}$~&~1.441~&~1.568~&~1.717~&~1.908~&~2.141~&~2.439~&~2.848~&~3.416~&~4.222~&~5.239~&~$10^{-8}$\\
 2.512$\times 10^{3}$~&~0.643~&~0.702~&~0.775~&~0.861~&~0.973~&~1.119~&~1.318~&~1.597~&~2.007~&~2.511~&~$10^{-8}$\\
 3.162$\times 10^{3}$~&~0.285~&~0.312~&~0.346~&~0.387~&~0.438~&~0.508~&~0.605~&~0.742~&~0.945~&~1.197~&~$10^{-8}$\\
 3.981$\times 10^{3}$~&~1.251~&~1.375~&~1.530~&~1.724~&~1.965~&~2.286~&~2.757~&~3.422~&~4.400~&~5.675~&~$10^{-9}$\\
 5.012$\times 10^{3}$~&~0.548~&~0.602~&~0.675~&~0.759~&~0.878~&~1.024~&~1.243~&~1.553~&~2.047~&~2.670~&~$10^{-9}$\\
 6.310$\times 10^{3}$~&~0.238~&~0.264~&~0.296~&~0.335~&~0.389~&~0.457~&~0.556~&~0.706~&~0.944~&~1.237~&~$10^{-9}$\\
 7.943$\times 10^{3}$~&~1.032~&~1.156~&~1.284~&~1.473~&~1.694~&~2.021~&~2.466~&~3.196~&~4.304~&~5.676~&~$10^{-10}$\\
 1.000$\times 10^{4}$~&~0.444~&~0.497~&~0.556~&~0.635~&~0.732~&~0.882~&~1.079~&~1.410~&~1.946~&~2.615~&~$10^{-10}$\\
\botrule
\end{supertabular}

\begin{table}[tbh]
\caption{$\bar\nu_\mu$ flux ($\rm m^{-2}sec^{-1}sr^{-1}GeV^{-1}$) above 10~GeV}
\end{table} 
\begin{supertabular}{c | r r r r r r r r r r | c }% \hline
\toprule
\multicolumn{1}{c|}{}&\multicolumn{10}{c}{$\cos\theta_z$}&\multicolumn{1}{|c}{}\\
$E_\nu$~(GeV)~&~1.--.9~&~.9--.8~&~.8--.7~&~.7--.6~&~.6--.5~&~.5--.4~&~.4--.3~&~.3--.2~&~.2--.1~&~.1--.0~&$Norm$\\
\toprule
 1.000$\times 10^{1}$~&~1.921~&~1.997~&~2.085~&~2.196~&~2.329~&~2.501~&~2.725~&~3.038~&~3.480~&~3.925~&~$10^{-1}$\\
 1.259$\times 10^{1}$~&~0.965~&~1.002~&~1.046~&~1.103~&~1.169~&~1.255~&~1.369~&~1.531~&~1.768~&~2.030~&~$10^{-1}$\\
 1.585$\times 10^{1}$~&~0.484~&~0.503~&~0.524~&~0.553~&~0.585~&~0.630~&~0.688~&~0.769~&~0.898~&~1.047~&~$10^{-1}$\\
 1.995$\times 10^{1}$~&~2.419~&~2.521~&~2.628~&~2.774~&~2.936~&~3.157~&~3.445~&~3.869~&~4.526~&~5.334~&~$10^{-2}$\\
 2.512$\times 10^{1}$~&~1.206~&~1.257~&~1.313~&~1.388~&~1.470~&~1.578~&~1.726~&~1.939~&~2.276~&~2.706~&~$10^{-2}$\\
 3.162$\times 10^{1}$~&~0.599~&~0.624~&~0.654~&~0.692~&~0.734~&~0.788~&~0.866~&~0.970~&~1.144~&~1.378~&~$10^{-2}$\\
 3.981$\times 10^{1}$~&~2.960~&~3.092~&~3.253~&~3.433~&~3.667~&~3.934~&~4.341~&~4.876~&~5.742~&~7.037~&~$10^{-3}$\\
 5.012$\times 10^{1}$~&~1.456~&~1.523~&~1.603~&~1.699~&~1.815~&~1.958~&~2.157~&~2.430~&~2.869~&~3.549~&~$10^{-3}$\\
 6.310$\times 10^{1}$~&~0.713~&~0.746~&~0.784~&~0.836~&~0.892~&~0.970~&~1.067~&~1.203~&~1.431~&~1.768~&~$10^{-3}$\\
 7.943$\times 10^{1}$~&~3.472~&~3.653~&~3.847~&~4.076~&~4.393~&~4.787~&~5.296~&~5.998~&~7.149~&~8.824~&~$10^{-4}$\\
 1.000$\times 10^{2}$~&~1.685~&~1.777~&~1.872~&~1.987~&~2.151~&~2.349~&~2.610~&~2.971~&~3.547~&~4.404~&~$10^{-4}$\\
 1.259$\times 10^{2}$~&~0.814~&~0.857~&~0.903~&~0.971~&~1.044~&~1.144~&~1.273~&~1.459~&~1.746~&~2.184~&~$10^{-4}$\\
 1.585$\times 10^{2}$~&~0.390~&~0.412~&~0.437~&~0.470~&~0.506~&~0.555~&~0.620~&~0.717~&~0.859~&~1.071~&~$10^{-4}$\\
 1.995$\times 10^{2}$~&~1.875~&~1.976~&~2.104~&~2.259~&~2.444~&~2.684~&~3.006~&~3.503~&~4.197~&~5.235~&~$10^{-5}$\\
 2.512$\times 10^{2}$~&~0.898~&~0.943~&~1.007~&~1.083~&~1.173~&~1.296~&~1.450~&~1.692~&~2.037~&~2.561~&~$10^{-5}$\\
 3.162$\times 10^{2}$~&~0.425~&~0.449~&~0.482~&~0.518~&~0.562~&~0.622~&~0.699~&~0.814~&~0.992~&~1.247~&~$10^{-5}$\\
 3.981$\times 10^{2}$~&~2.007~&~2.136~&~2.288~&~2.469~&~2.691~&~2.975~&~3.364~&~3.928~&~4.804~&~6.034~&~$10^{-6}$\\
 5.012$\times 10^{2}$~&~0.945~&~1.011~&~1.078~&~1.169~&~1.283~&~1.422~&~1.613~&~1.893~&~2.311~&~2.908~&~$10^{-6}$\\
 6.310$\times 10^{2}$~&~0.441~&~0.472~&~0.508~&~0.551~&~0.608~&~0.675~&~0.770~&~0.903~&~1.110~&~1.397~&~$10^{-6}$\\
 7.943$\times 10^{2}$~&~2.045~&~2.189~&~2.374~&~2.578~&~2.855~&~3.190~&~3.651~&~4.310~&~5.324~&~6.682~&~$10^{-7}$\\
 1.000$\times 10^{3}$~&~0.941~&~1.012~&~1.096~&~1.198~&~1.331~&~1.500~&~1.724~&~2.054~&~2.543~&~3.181~&~$10^{-7}$\\
 1.259$\times 10^{3}$~&~0.428~&~0.463~&~0.505~&~0.555~&~0.617~&~0.700~&~0.811~&~0.968~&~1.208~&~1.514~&~$10^{-7}$\\
 1.585$\times 10^{3}$~&~1.930~&~2.095~&~2.298~&~2.534~&~2.839~&~3.244~&~3.777~&~4.548~&~5.717~&~7.207~&~$10^{-8}$\\
 1.995$\times 10^{3}$~&~0.863~&~0.939~&~1.032~&~1.143~&~1.295~&~1.489~&~1.743~&~2.129~&~2.692~&~3.416~&~$10^{-8}$\\
 2.512$\times 10^{3}$~&~0.379~&~0.418~&~0.461~&~0.515~&~0.584~&~0.675~&~0.799~&~0.983~&~1.260~&~1.602~&~$10^{-8}$\\
 3.162$\times 10^{3}$~&~1.672~&~1.846~&~2.046~&~2.293~&~2.610~&~3.044~&~3.641~&~4.523~&~5.860~&~7.522~&~$10^{-9}$\\
 3.981$\times 10^{3}$~&~0.736~&~0.810~&~0.900~&~1.014~&~1.158~&~1.361~&~1.646~&~2.069~&~2.704~&~3.526~&~$10^{-9}$\\
 5.012$\times 10^{3}$~&~0.316~&~0.352~&~0.394~&~0.449~&~0.510~&~0.600~&~0.734~&~0.931~&~1.238~&~1.627~&~$10^{-9}$\\
 6.310$\times 10^{3}$~&~1.363~&~1.534~&~1.714~&~1.946~&~2.232~&~2.662~&~3.225~&~4.149~&~5.645~&~7.456~&~$10^{-10}$\\
 7.943$\times 10^{3}$~&~0.593~&~0.665~&~0.741~&~0.832~&~0.974~&~1.185~&~1.417~&~1.846~&~2.556~&~3.404~&~$10^{-10}$\\
 1.000$\times 10^{4}$~&~0.256~&~0.281~&~0.318~&~0.361~&~0.424~&~0.516~&~0.634~&~0.820~&~1.140~&~1.541~&~$10^{-10}$\\
\botrule
\end{supertabular}
\clearpage

%\vspace{1cm}
\begin{table}[tbh]
\caption{$\nu_e$ flux ($\rm m^{-2}sec^{-1}sr^{-1}GeV^{-1}$) above 10~GeV}
\end{table} 
\begin{supertabular}{c | r r r r r r r r r r | c }% \hline
\toprule
\multicolumn{1}{c|}{}&\multicolumn{10}{c}{$\cos\theta_z$}&\multicolumn{1}{|c}{}\\
$E_\nu$~(GeV)~&~1.--.9~&~.9--.8~&~.8--.7~&~.7--.6~&~.6--.5~&~.5--.4~&~.4--.3~&~.3--.2~&~.2--.1~&~.1--.0~~&~$Norm$\\
\toprule
 1.000$\times 10^{1}$~&~0.431~&~0.484~&~0.548~&~0.634~&~0.742~&~0.888~&~1.090~&~1.383~&~1.798~&~2.242~&~$10^{-1}$\\
 1.259$\times 10^{1}$~&~0.191~&~0.215~&~0.245~&~0.284~&~0.335~&~0.404~&~0.503~&~0.650~&~0.871~&~1.131~&~$10^{-1}$\\
 1.585$\times 10^{1}$~&~0.848~&~0.951~&~1.088~&~1.263~&~1.497~&~1.819~&~2.296~&~3.015~&~4.177~&~5.624~&~$10^{-2}$\\
 1.995$\times 10^{1}$~&~0.374~&~0.421~&~0.479~&~0.561~&~0.665~&~0.817~&~1.033~&~1.386~&~1.968~&~2.759~&~$10^{-2}$\\
 2.512$\times 10^{1}$~&~0.166~&~0.187~&~0.212~&~0.248~&~0.295~&~0.363~&~0.463~&~0.630~&~0.917~&~1.335~&~$10^{-2}$\\
 3.162$\times 10^{1}$~&~0.740~&~0.831~&~0.942~&~1.102~&~1.308~&~1.604~&~2.073~&~2.841~&~4.240~&~6.414~&~$10^{-3}$\\
 3.981$\times 10^{1}$~&~0.331~&~0.371~&~0.419~&~0.489~&~0.580~&~0.714~&~0.926~&~1.284~&~1.945~&~3.080~&~$10^{-3}$\\
 5.012$\times 10^{1}$~&~0.151~&~0.167~&~0.187~&~0.216~&~0.256~&~0.316~&~0.410~&~0.570~&~0.881~&~1.451~&~$10^{-3}$\\
 6.310$\times 10^{1}$~&~0.685~&~0.752~&~0.842~&~0.959~&~1.130~&~1.390~&~1.804~&~2.501~&~3.952~&~6.684~&~$10^{-4}$\\
 7.943$\times 10^{1}$~&~0.303~&~0.340~&~0.373~&~0.429~&~0.504~&~0.611~&~0.792~&~1.110~&~1.770~&~3.062~&~$10^{-4}$\\
 1.000$\times 10^{2}$~&~0.138~&~0.154~&~0.169~&~0.194~&~0.226~&~0.272~&~0.350~&~0.488~&~0.781~&~1.388~&~$10^{-4}$\\
 1.259$\times 10^{2}$~&~0.649~&~0.704~&~0.771~&~0.883~&~1.018~&~1.219~&~1.552~&~2.122~&~3.391~&~6.202~&~$10^{-5}$\\
 1.585$\times 10^{2}$~&~0.290~&~0.323~&~0.347~&~0.395~&~0.461~&~0.546~&~0.686~&~0.938~&~1.480~&~2.740~&~$10^{-5}$\\
 1.995$\times 10^{2}$~&~0.132~&~0.148~&~0.161~&~0.180~&~0.209~&~0.245~&~0.305~&~0.414~&~0.649~&~1.203~&~$10^{-5}$\\
 2.512$\times 10^{2}$~&~0.617~&~0.673~&~0.758~&~0.828~&~0.947~&~1.103~&~1.367~&~1.812~&~2.853~&~5.262~&~$10^{-6}$\\
 3.162$\times 10^{2}$~&~0.285~&~0.308~&~0.344~&~0.377~&~0.431~&~0.500~&~0.616~&~0.803~&~1.249~&~2.283~&~$10^{-6}$\\
 3.981$\times 10^{2}$~&~1.299~&~1.406~&~1.541~&~1.714~&~1.957~&~2.275~&~2.767~&~3.600~&~5.457~&~9.861~&~$10^{-7}$\\
 5.012$\times 10^{2}$~&~0.591~&~0.640~&~0.700~&~0.782~&~0.888~&~1.033~&~1.238~&~1.613~&~2.387~&~4.247~&~$10^{-7}$\\
 6.310$\times 10^{2}$~&~0.268~&~0.289~&~0.320~&~0.354~&~0.406~&~0.467~&~0.555~&~0.715~&~1.049~&~1.825~&~$10^{-7}$\\
 7.943$\times 10^{2}$~&~1.204~&~1.298~&~1.441~&~1.609~&~1.836~&~2.132~&~2.532~&~3.251~&~4.645~&~7.852~&~$10^{-8}$\\
 1.000$\times 10^{3}$~&~0.536~&~0.583~&~0.639~&~0.732~&~0.814~&~0.973~&~1.161~&~1.498~&~2.076~&~3.387~&~$10^{-8}$\\
 1.259$\times 10^{3}$~&~0.237~&~0.262~&~0.285~&~0.327~&~0.361~&~0.435~&~0.521~&~0.667~&~0.934~&~1.464~&~$10^{-8}$\\
 1.585$\times 10^{3}$~&~1.037~&~1.155~&~1.268~&~1.427~&~1.652~&~1.935~&~2.327~&~3.011~&~4.176~&~6.411~&~$10^{-9}$\\
 1.995$\times 10^{3}$~&~0.453~&~0.501~&~0.560~&~0.624~&~0.752~&~0.859~&~1.042~&~1.365~&~1.859~&~2.816~&~$10^{-9}$\\
 2.512$\times 10^{3}$~&~0.199~&~0.217~&~0.248~&~0.281~&~0.319~&~0.378~&~0.463~&~0.588~&~0.830~&~1.217~&~$10^{-9}$\\
 3.162$\times 10^{3}$~&~0.872~&~0.954~&~1.073~&~1.216~&~1.382~&~1.651~&~2.051~&~2.620~&~3.683~&~5.416~&~$10^{-10}$\\
 3.981$\times 10^{3}$~&~0.377~&~0.422~&~0.458~&~0.517~&~0.626~&~0.722~&~0.897~&~1.199~&~1.647~&~2.483~&~$10^{-10}$\\
 5.012$\times 10^{3}$~&~0.164~&~0.183~&~0.201~&~0.236~&~0.275~&~0.321~&~0.383~&~0.518~&~0.759~&~1.114~&~$10^{-10}$\\
 6.310$\times 10^{3}$~&~0.716~&~0.781~&~0.843~&~1.020~&~1.180~&~1.465~&~1.692~&~2.282~&~3.355~&~4.746~&~$10^{-11}$\\
 7.943$\times 10^{3}$~&~0.305~&~0.333~&~0.353~&~0.415~&~0.505~&~0.647~&~0.753~&~0.993~&~1.434~&~2.034~&~$10^{-11}$\\
 1.000$\times 10^{4}$~&~1.233~&~1.428~&~1.637~&~1.773~&~2.170~&~2.604~&~3.143~&~3.900~&~6.388~&~9.422~&~$10^{-12}$\\
\botrule
\end{supertabular}

\begin{table}[tbh]
\caption{$\bar\nu_e$ flux ($\rm m^{-2}sec^{-1}sr^{-1}GeV^{-1}$) above 10~GeV}
\end{table} 
\begin{supertabular}{c | r r r r r r r r r r | c }% \hline
\toprule
\multicolumn{1}{c|}{}&\multicolumn{10}{c}{$\cos\theta_z$}&\multicolumn{1}{|c}{}\\
$E_\nu$~(GeV)~&~1.--.9~&~.9--.8~&~.8--.7~&~.7--.6~&~.6--.5~&~.5--.4~&~.4--.3~&~.3--.2~&~.2--.1~&~.1--.0~~&~$Norm$\\
\toprule
 1.000$\times 10^{1}$~&~0.327~&~0.367~&~0.416~&~0.478~&~0.560~&~0.668~&~0.819~&~1.038~&~1.352~&~1.687~&~$10^{-1}$\\
 1.259$\times 10^{1}$~&~1.465~&~1.639~&~1.858~&~2.153~&~2.531~&~3.051~&~3.779~&~4.880~&~6.539~&~8.490~&~$10^{-2}$\\
 1.585$\times 10^{1}$~&~0.654~&~0.733~&~0.830~&~0.962~&~1.134~&~1.379~&~1.731~&~2.269~&~3.153~&~4.241~&~$10^{-2}$\\
 1.995$\times 10^{1}$~&~0.291~&~0.327~&~0.372~&~0.431~&~0.509~&~0.622~&~0.787~&~1.047~&~1.494~&~2.081~&~$10^{-2}$\\
 2.512$\times 10^{1}$~&~0.130~&~0.146~&~0.167~&~0.192~&~0.228~&~0.279~&~0.355~&~0.478~&~0.698~&~1.010~&~$10^{-2}$\\
 3.162$\times 10^{1}$~&~0.589~&~0.652~&~0.746~&~0.855~&~1.021~&~1.251~&~1.599~&~2.175~&~3.239~&~4.890~&~$10^{-3}$\\
 3.981$\times 10^{1}$~&~0.267~&~0.296~&~0.337~&~0.386~&~0.457~&~0.562~&~0.721~&~0.990~&~1.496~&~2.362~&~$10^{-3}$\\
 5.012$\times 10^{1}$~&~0.121~&~0.134~&~0.152~&~0.173~&~0.204~&~0.250~&~0.323~&~0.446~&~0.681~&~1.116~&~$10^{-3}$\\
 6.310$\times 10^{1}$~&~0.550~&~0.602~&~0.684~&~0.769~&~0.910~&~1.104~&~1.432~&~1.978~&~3.057~&~5.166~&~$10^{-4}$\\
 7.943$\times 10^{1}$~&~0.250~&~0.277~&~0.311~&~0.344~&~0.404~&~0.493~&~0.633~&~0.871~&~1.364~&~2.391~&~$10^{-4}$\\
 1.000$\times 10^{2}$~&~0.115~&~0.126~&~0.141~&~0.156~&~0.181~&~0.220~&~0.283~&~0.385~&~0.605~&~1.086~&~$10^{-4}$\\
 1.259$\times 10^{2}$~&~0.536~&~0.570~&~0.634~&~0.717~&~0.820~&~0.985~&~1.267~&~1.706~&~2.678~&~4.847~&~$10^{-5}$\\
 1.585$\times 10^{2}$~&~0.245~&~0.261~&~0.288~&~0.324~&~0.374~&~0.441~&~0.556~&~0.749~&~1.186~&~2.162~&~$10^{-5}$\\
 1.995$\times 10^{2}$~&~1.113~&~1.209~&~1.307~&~1.454~&~1.686~&~1.981~&~2.448~&~3.312~&~5.199~&~9.476~&~$10^{-6}$\\
 2.512$\times 10^{2}$~&~0.505~&~0.555~&~0.596~&~0.656~&~0.756~&~0.893~&~1.100~&~1.481~&~2.264~&~4.107~&~$10^{-6}$\\
 3.162$\times 10^{2}$~&~0.228~&~0.249~&~0.274~&~0.302~&~0.346~&~0.404~&~0.500~&~0.660~&~0.995~&~1.806~&~$10^{-6}$\\
 3.981$\times 10^{2}$~&~1.033~&~1.129~&~1.246~&~1.385~&~1.583~&~1.836~&~2.246~&~2.946~&~4.378~&~7.861~&~$10^{-7}$\\
 5.012$\times 10^{2}$~&~0.470~&~0.514~&~0.559~&~0.625~&~0.716~&~0.832~&~1.006~&~1.318~&~1.926~&~3.373~&~$10^{-7}$\\
 6.310$\times 10^{2}$~&~0.212~&~0.228~&~0.252~&~0.280~&~0.321~&~0.372~&~0.458~&~0.590~&~0.856~&~1.458~&~$10^{-7}$\\
 7.943$\times 10^{2}$~&~0.936~&~1.011~&~1.133~&~1.256~&~1.441~&~1.660~&~2.046~&~2.697~&~3.818~&~6.328~&~$10^{-8}$\\
 1.000$\times 10^{3}$~&~0.408~&~0.451~&~0.503~&~0.562~&~0.646~&~0.746~&~0.904~&~1.227~&~1.697~&~2.738~&~$10^{-8}$\\
 1.259$\times 10^{3}$~&~0.180~&~0.200~&~0.221~&~0.248~&~0.287~&~0.337~&~0.412~&~0.529~&~0.748~&~1.178~&~$10^{-8}$\\
 1.585$\times 10^{3}$~&~0.796~&~0.870~&~0.967~&~1.099~&~1.272~&~1.508~&~1.846~&~2.330~&~3.312~&~5.101~&~$10^{-9}$\\
 1.995$\times 10^{3}$~&~0.347~&~0.379~&~0.423~&~0.485~&~0.566~&~0.659~&~0.806~&~1.054~&~1.473~&~2.236~&~$10^{-9}$\\
 2.512$\times 10^{3}$~&~1.499~&~1.665~&~1.824~&~2.101~&~2.514~&~2.847~&~3.542~&~4.578~&~6.517~&~9.874~&~$10^{-10}$\\
 3.162$\times 10^{3}$~&~0.643~&~0.717~&~0.774~&~0.898~&~1.086~&~1.259~&~1.537~&~2.057~&~2.870~&~4.396~&~$10^{-10}$\\
 3.981$\times 10^{3}$~&~0.277~&~0.305~&~0.330~&~0.388~&~0.456~&~0.557~&~0.663~&~0.915~&~1.270~&~1.919~&~$10^{-10}$\\
 5.012$\times 10^{3}$~&~1.197~&~1.328~&~1.434~&~1.723~&~1.928~&~2.333~&~2.939~&~3.614~&~5.617~&~7.958~&~$10^{-11}$\\
 6.310$\times 10^{3}$~&~0.507~&~0.557~&~0.618~&~0.726~&~0.859~&~1.003~&~1.299~&~1.576~&~2.278~&~3.389~&~$10^{-11}$\\
 7.943$\times 10^{3}$~&~0.212~&~0.227~&~0.263~&~0.305~&~0.381~&~0.453~&~0.551~&~0.739~&~0.932~&~1.522~&~$10^{-11}$\\
 1.000$\times 10^{4}$~&~0.910~&~0.973~&~1.133~&~1.435~&~1.529~&~1.997~&~2.215~&~2.971~&~4.573~&~6.825~&~$10^{-12}$\\
\botrule
\end{supertabular}

\end{document}